\documentclass[reprint,aps,prl,superscriptaddress,longbibliography]{revtex4-1}
\usepackage[latin9]{inputenc}
\usepackage{times}
\usepackage{color}
\usepackage{amsmath}
\usepackage{amssymb}
\usepackage{stmaryrd}
\usepackage{bbm}
\usepackage{graphicx}
\usepackage{soul}
\usepackage[unicode=true,
 bookmarks=true,bookmarksnumbered=false,bookmarksopen=false,
 breaklinks=false,pdfborder={0 0 1},backref=false,colorlinks=true]
 {hyperref}
\hypersetup{
 linkcolor=magenta, urlcolor=blue, citecolor=blue, pdfstartview={FitH}, hyperfootnotes=true, unicode=true}
 
\makeatletter
\@ifundefined{textcolor}{}
{%
 \definecolor{BLACK}{gray}{0}
 \definecolor{WHITE}{gray}{1}
 \definecolor{RED}{rgb}{1,0,0}
 \definecolor{GREEN}{rgb}{0,1,0}
 \definecolor{BLUE}{rgb}{0,0,1}
 \definecolor{CYAN}{cmyk}{1,0,0,0}
 \definecolor{MAGENTA}{cmyk}{0,1,0,0}
 \definecolor{YELLOW}{cmyk}{0,0,1,0}
}

\newcommand{\doublewidetilde}[1]{{%
  \mathpalette\double@widetilde{#1}%
}}
\newcommand{\double@widetilde}[2]{%
  \sbox\z@{$\m@th#1\widetilde{#2}$}%
  \ht\z@=.9\ht\z@
  \widetilde{\box\z@}%
}

\usepackage{amsfonts}\usepackage{tabularx}\usepackage{dcolumn}\usepackage{bm}\usepackage{graphicx}\usepackage{epstopdf}

\hypersetup{urlcolor=blue}

\usepackage{tensor}
\usepackage{braket}

\newcommand{\Hc}{\mathrm{H.c.}}
\DeclareMathOperator{\tr}{tr}

\newcommand{\beq}{\begin{equation}}
\newcommand{\eeq}{\end{equation}}
\newcommand{\nn}{\nonumber \\[.25cm]}
\def\bea{\begin{eqnarray}}
\def\eea{\end{eqnarray}}

\usepackage[capitalise,compress]{cleveref}
\crefname{section}{Sec.}{Secs.}
\Crefname{section}{Section}{Sections}
\crefrangelabelformat{equation}{\textup{(#3#1#4)}--\textup{(#5#2#6)}}

\makeatother
\begin{document}

\title{Circuit Complexity across a Topological Phase Transition
}
\author{Fangli Liu}
\affiliation{Joint Quantum Institute, NIST/University of Maryland, College Park, Maryland 20742, USA}
\affiliation{Joint Center for Quantum Information and Computer Science, NIST/University of Maryland, College Park, MD 20742, USA}
\author{Seth Whitsitt}
\affiliation{Joint Quantum Institute, NIST/University of Maryland, College Park, Maryland 20742, USA}
\affiliation{Joint Center for Quantum Information and Computer Science, NIST/University of Maryland, College Park, MD 20742, USA}
\author{Jonathan B. Curtis}
\affiliation{Joint Quantum Institute, NIST/University of Maryland, College Park, Maryland 20742, USA}
\author{Rex Lundgren}
\affiliation{Joint Quantum Institute, NIST/University of Maryland, College Park, Maryland 20742, USA}
\author{Paraj Titum}
\affiliation{Joint Quantum Institute, NIST/University of Maryland, College Park, Maryland 20742, USA}
\affiliation{Joint Center for Quantum Information and Computer Science, NIST/University of Maryland, College Park, MD 20742, USA}

\author{Zhi-Cheng Yang}
\affiliation{Joint Quantum Institute, NIST/University of Maryland, College Park, Maryland 20742, USA}
\affiliation{Joint Center for Quantum Information and Computer Science, NIST/University of Maryland, College Park, MD 20742, USA}
\author{James R. Garrison}
\affiliation{Joint Quantum Institute, NIST/University of Maryland, College Park, Maryland 20742, USA}
\affiliation{Joint Center for Quantum Information and Computer Science, NIST/University of Maryland, College Park, MD 20742, USA}
\author{Alexey V. Gorshkov}
\affiliation{Joint Quantum Institute, NIST/University of Maryland, College Park, Maryland 20742, USA}
\affiliation{Joint Center for Quantum Information and Computer Science, NIST/University of Maryland, College Park, MD 20742, USA}
\begin{abstract}
We use Nielsen's geometric approach to quantify the circuit complexity in a one-dimensional Kitaev chain across a topological phase transition. We find that the circuit complexities of both the ground states and non-equilibrium steady states of the Kitaev model exhibit non-analytical behaviors at the critical points, and thus can be used to detect both {\it equilibrium} and {\it dynamical} topological phase transitions. Moreover, we show that the locality property of the real-space optimal Hamiltonian connecting two different ground states depends crucially on whether the two states belong to the same or different phases. This provides a concrete example of classifying different gapped phases using Nielsen's circuit complexity.  We further generalize our results to a Kitaev chain with long-range pairing, and discuss generalizations to higher dimensions. Our result opens up a new avenue for using circuit complexity as a novel tool to understand quantum many-body systems.

\end{abstract}


\maketitle
In computer science, the notion of computational complexity refers to the minimum number of elementary operations for  implementing a given task. This concept readily extends to quantum information science, where quantum circuit complexity denotes the minimum number of gates  to implement a desired unitary transformation. The corresponding circuit complexity of a  quantum state  characterizes how difficult it is to  construct a unitary transformation $U$  which evolves a reference state  to the desired target state \cite{Watrous2009, Aaronson16}.  Nielsen and collaborators used a geometric approach to tackle the problem of quantum complexity \cite{Nielsen05, Nielsen06, Dowling07}. Suppose that the unitary transformation $U(t)$ is generated by some time-dependent Hamiltonian $H(t)$, with the requirement that $U(t_f)=U$ (where $t_f$ denotes the final time). Then, the quantum state complexity is quantified by imposing a cost functional $F[H(t)]$ on the control Hamiltonian $H(t)$.  By choosing a cost functional that defines a Riemannian geometry in the space of circuits, the problem of finding the optimal control Hamiltonian synthesizing $U$ then corresponds to finding minimal geodesic paths in a Riemannian geometry  \cite{Nielsen05, Nielsen06, Dowling07}. 

Recently, Nielsen's approach has been adopted in high-energy physics  to quantify the complexity of  quantum field theory states \cite{Jefferson17, Yang18, Guo18, Camargo18, Alves18, Chapman18, Hackl18, Khan18, Rey17, Jiang18, YRQ19_1, YRQ19_2, YRQ19_3}.  This is motivated, in part, by previous conjectures that relate the complexity of the boundary field theory to the bulk space-time geometry, i.e.\ the so-called ``complexity equals volume" \cite{Stanford14, Susskind14} and  ``complexity equals action''  \cite{Brown15, Brown16} proposals.   Jefferson {\it et al.}\ used Nielsen's  approach to calculate the complexity of a free scalar field \cite{Jefferson17}, and found surprising similarities to the results of holographic complexity. A complementary study by Chapman et al., using the Fubini-Study metric to quantify complexity \cite{Chapman18PRL}, gave similar results. Several recent works have generalized these studies to other states, including coherent states \cite{Guo18, Caputa18}, thermofield double states \cite{Chapman18, Yang18},  and free fermion fields \cite{Hackl18, Khan18, Rey17}. However, the connection between the geometric definition of circuit complexity and quantum phase transitions has so far remained unexplored. This connection is important both fundamentally, and is also intimately related to the long-standing problem of quantum state preparations across critical points \cite{Vojta03,Caneva, Sorensen10}.

In this work, we consider the circuit complexity of a  topological quantum system.
In particular, we use Nielsen's approach to study the circuit complexity of the Kitaev chain, a prototypical model exhibiting topological phase transitions and hosting Majorana zero modes \cite{Kitaev01, Alicea12, Jason11, Sau10, Oreg10, Lutch10}.  Strikingly, we find that the circuit complexity derived using this approach exhibits non-analytical behaviors at the critical points, for both {\it equilibrium} and {\it dynamical} topological phase transitions. Moreover, the optimal Hamiltonian connecting the initial and final states must be non-local in {\it real-space} when evolving across a critical point. We further generalize our results to a Kitaev chain with long-range pairing, and discuss universal features of non-analyticities at the critical points in higher dimensions. Our work establishes a connection between geometrical circuit complexity and quantum phase transitions, and paves the way towards using complexity as a novel tool to study quantum many-body systems.

\begin{figure}
  \centering\includegraphics[width=0.49\textwidth, height=7.5cm]{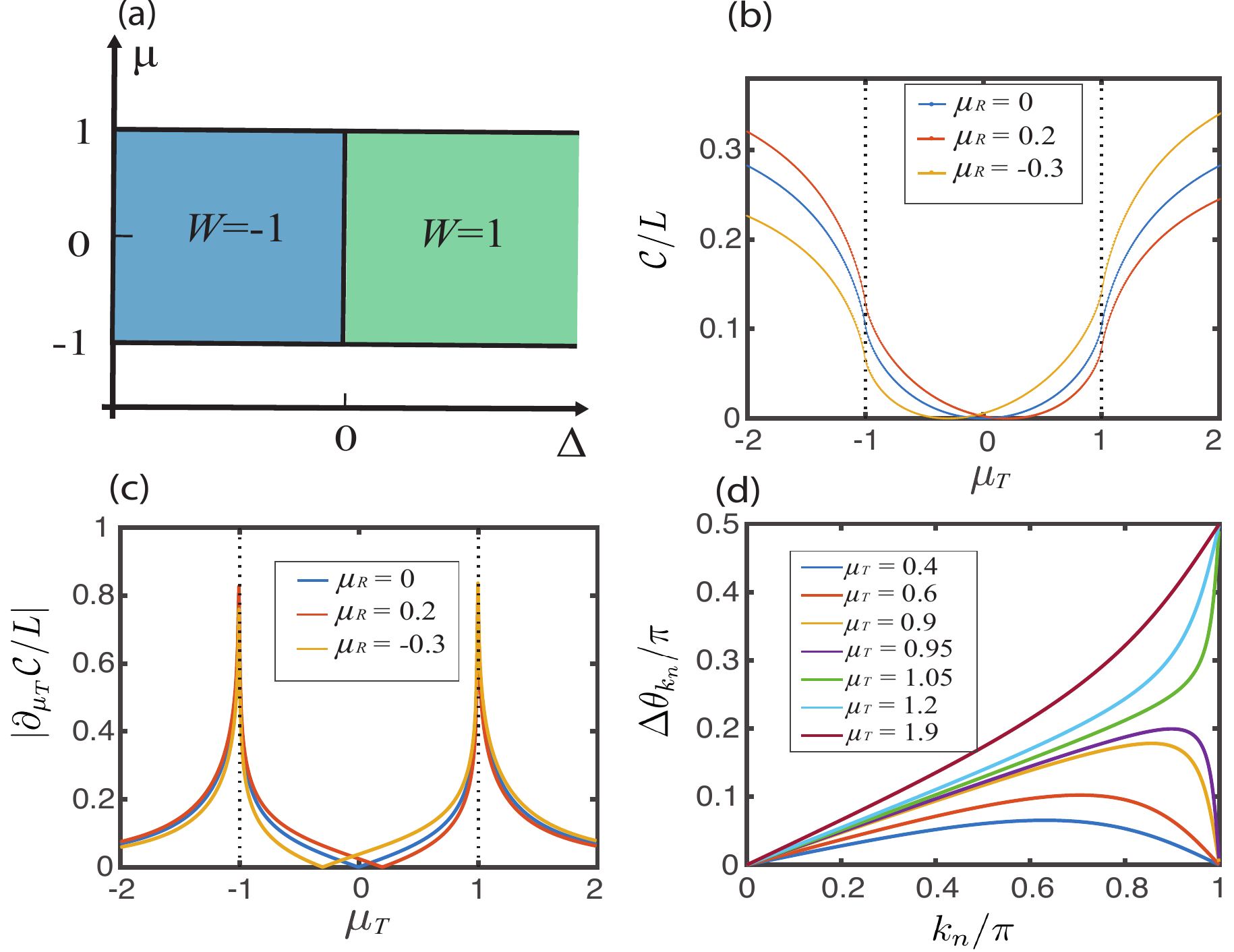}
  \caption{(a) Phase diagram of the Kitaev chain, with $W$ denoting the winding number.  (b) Ground state circuit complexity and (c) its derivative versus target state chemical potential ($\mu_T$) for several reference states, each with a different chemical potential $\mu_R$.  (d) Bogoliubov angle difference, $\Delta \theta_{k_n}$, for different target ground states, with $\mu_R= 0$.  $\Delta_{R}=\Delta_{T}=1$ for (b)--(d), and $L=1000$ for (b) and (c).
  }
  \label{fig1}
\end{figure}

{\it The model.---}%
The 1D Kitaev model is described by the following Hamiltonian \cite{Kitaev01, Alicea12}:
\begin{equation}
\begin{split}
\hat{H}=&- \frac{J}{2} \sum_{j=1}^{L}\left( \hat{a}_j^{\dagger}\hat{a}^{\phantom\dagger}_{j+1} +\Hc\right) -\mu \sum_{j=1}^{L}\left(\hat{a}_j^{\dagger} \hat{a}^{\phantom\dagger}_j -\frac{1}{2}\right) \\  &+ \frac{\Delta}{2} \sum_{j=1}^{L}\left(\hat{a}_j^{\dagger} \hat{a}_{j+1}^{\dagger} +\Hc\right), 
\end{split}
\label{kitaev}
\end{equation}
where $J$ is the hopping amplitude, $\Delta$ is the superconducting pairing strength, $\mu$ is the chemical potential,  $L$ is the total number of sites (assumed to be even), and  $\hat{a}_j^{\dagger}$ ($\hat{a}^{\phantom\dagger}_j$) creates (annihilates) a fermion at site $j$.  We set $J=1$  and assume antiperiodic boundary conditions ($\hat{a}_{L+1}=-\hat{a}_1$).
Upon Fourier transforming  \cref{kitaev} can be written in the momentum basis
\begin{equation}
\begin{split}
\hat{H}= -\sum_{k_n}& [\mu+\cos k_n] \left( \hat{a}_{k_n}^{\dagger} \hat{a}^{\phantom\dagger}_{k_n} - \hat{a}^{\phantom\dagger}_{-k_n} \hat{a}_{-k_n}^{\dagger}\right) \\  & + i \Delta \sin k_n \left(\hat{a}_{k_n}^{\dagger}\hat{a}_{-k_n}^{\dagger}- \hat{a}^{\phantom\dagger}_{-k_n} \hat{a}^{\phantom\dagger}_{k_n}\right), 
\end{split}
\label{k_space}
\end{equation}
where $k_n=\frac{2\pi}{L}(n+1/2)$  with $n= 0, 1, \dots, L/2-1$. The above Hamiltonian can be diagonalized  via a Bogoliubov transformation, which yields the   excitation spectrum: $\varepsilon_{k_n}= \sqrt{(\mu+\cos k_n )^2+ \Delta^2\sin^2k_n}. $ The ground state of Eq.~\eqref{kitaev} can be written as 
\begin{equation}
\ket{\Psi_{\text{gs}}}= \prod_{n=0}^{L/2-1} (\cos \theta_{k_n} - i\sin \theta_{k_n} \hat{a}_{k_n}^{\dagger}\hat{a}_{-k_n}^{\dagger}) \ket{0},
\label{ground}
\end{equation}
where $\tan(2 \theta_{k_n})= \Delta \sin k_n/ (\mu + \cos k_n)$.  A topological phase transition occurs when the quasiparticle spectrum is gapless \cite{Kitaev01}, as illustrated in Fig.~\ref{fig1}(a). The nontrivial topological phase is characterized by a nonzero winding number and the presence of  Majorana edge modes  \cite{Kitaev01, Alicea12, Jason11, Sau10, Oreg10, Lutch10}.  


{\it Complexity for a pair of fermions.---}%
Since Hamiltonian~(\ref{kitaev}) is non-interacting, the ground state wavefunction~(\ref{ground}) couples only pairs of fermionic modes with momenta $\pm k_n$, and different momentum pairs are decoupled. Hence, we first compute the circuit complexity of one such fermionic pair \cite{Hackl18, Khan18, Rey17}, and then obtain the complexity of the full system by summing over all momentum contributions \cite{Jefferson17, Chapman18PRL}. 

Let us consider the reference (``$R$'') and target (``$T$'') states with the same momentum but different Bogoliubov angles: $\ket{\psi_{R,T} }= (\cos \theta^{R,T}_{k} - i\sin \theta^{R,T}_{k} \hat{a}_{k}^{\dagger}\hat{a}_{-k}^{\dagger}) \ket{0}$.  Expanding the target state in the  basis of  $\ket{\psi_{R} }$ and $\ket{\psi_R}_{\perp}$ (i.e., the state orthogonal to $\ket{\psi_R}$), we have $\ket{\psi_{T}}= \cos(\Delta \theta_{k}) \ket{\psi_R} - i\sin(\Delta \theta_{k}) \ket{\psi_R}_{\perp} $, where $\Delta \theta_{k}=  \theta^{R}_{k}-\theta^{T}_{k}$.  Now the goal is to find the optimal circuit to achieve the unitary transformation connecting $\ket{\psi_{R} }$ and $\ket{\psi_{T} }$ : 
\begin{equation}
U_k= \begin{bmatrix} 
    \cos(\Delta \theta_{k})& - i e^{-i\phi} \sin(\Delta \theta_{k})\\- i  \sin(\Delta \theta_{k})& e^{-i \phi}\cos(\Delta \theta_{k}) 
    \end{bmatrix},
    \label{unitary}
\end{equation}
where $\phi$ is an arbitrary phase. Nielsen approached this as a Hamiltonian control problem, i.e.\ finding a time-dependent Hamiltonian $\mathcal{H}_k(s) $ that synthesizes the trajectory in the space of unitaries~\cite{Nielsen05, Nielsen06}:
\begin{equation}
U_k(s)= \overleftarrow{\mathcal{P}} \exp \left[\int_0^{s} dt\, \mathcal{H}(t) \right],~\mathcal{H}_k(t)= \sum_IY_k^I(t) O_I
\label{u_k}
\end{equation}
with boundary conditions $U_k(s=0)= \mathbbm{1}$, and $U_k(s=1)= U_k$. Here, $\overleftarrow{\mathcal{P}}$ is the path-ordering operator  and $O_I$ are the generators of $U(2)$. The idea is then to define a {\it cost} (i.e.\ `length') functional for the various possible paths to achieve $U_k$ \cite{Nielsen05, Nielsen06, Jefferson17, Hackl18}: $\mathcal{D}\left[U_k\right]= \int_0^1 ds \sum_I |Y_k^I(s)|^2$, 
and to identify the optimal circuit or path by minimizing this functional. The  cost of the {\it optimal} path  is  called the circuit complexity $\mathcal{C}$ of the target state, i.e.\ 
\begin{equation}
\mathcal{C}\left[U_k\right] = \textrm{min}_{\left\{Y_k^I(s)\right\}} \mathcal{D}\left[U_k\right].
\end{equation}

Following the procedures in Refs.~\cite{Hackl18, Khan18, Rey17}, one can explicitly calculate the circuit complexity  for synthesizing the unitary transformation~(\ref{unitary}). For quadratic Hamiltonians, it is a simple expression that depends only on the difference between Bogoliubov angles (see Supplemental Material \cite{supp}),
\begin{equation}
\mathcal{C} \left( \ket{\psi_R} \rightarrow \ket{\psi_T} \right)  =|\Delta\theta_k| ^2.
\label{C2}
\end{equation}
Note that the complexity $\mathcal{C}$ for two fermions is at most $\pi^2/4$, since  $|\Delta\theta_{k}| \in [0, \pi/2]$. The maximum value is achieved when the target state has vanishing overlap with the reference state.

{\it Complexity for the full wavefunction.---}%
Given the circuit complexity for a pair of fermionic modes,  one can readily obtain the complexity of the full many-body wavefunction.  The total unitary transformation that connects the two different ground states [Eq.~\eqref{ground}] is:
\begin{equation}
\ket{\Psi_\text{gs}^T} = \left(\prod_{n=0}^{L/2-1} U_{k_n} \right) \ket{\Psi_\text{gs}^R},
\end{equation}
where $U_{k_n}$, given by Eq.~(\ref{unitary}), connects two fermionic states with momenta $\pm k_n$. By choosing the cost function  to be a summation of all momentum contributions \cite{Jefferson17, Hackl18, Khan18, Rey17}, it is straightforward to obtain the total circuit complexity 
\begin{equation}
  \mathcal{C}\left( \ket{\Psi_\text{gs}^R} \rightarrow \ket{\Psi_\text{gs}^{T}} \right) =\sum_{k_n} | \Delta \theta_{k_n} |^2,
  \label{comp}
\end{equation}
where $\Delta \theta_{k_n}$ is the difference of the Bogoliubov angles for momentum  $ k_n$.  In the infinite-system-size limit, the summation can be replaced by an integral, and one can derive that $\mathcal{C} \propto L$.  This ``volume law'' dependence is reminiscent of the ``complexity equals volume'' conjecture in holography \cite{Stanford14, Susskind14}, albeit in a different setting.

The circuit complexity given by Eq.\ (\ref{comp}) has a geometric interpretation, as it is the squared Euclidean distance in a high dimensional space \footnote{In such a space, each state is represented by one point, with its coordinates labeled by the Bogoliubov angles, i.e.\ $(\theta_{k_0}, \theta_{k_1}, \dots , \theta_{k_{L/2-1}})$ }. The geodesic path (or optimal circuit) in unitary space turns out to be a straight line connecting the two points [i.e.~$H_k(s)$ indepedent of $s$) \cite{supp}.
 In the remainder of this paper, we demonstrate that the circuit complexity between two states is  able to reveal both {\it equilibrium} and {\it dynamical} topological phase transitions.

 We first choose a fixed ground state as the reference state and calculate the circuit complexities for target ground states with various chemical potentials $\mu_T$, crossing the phase transition point.
The circuit complexity  increases as the difference between the  parameters of reference and target states  is increased [Fig.~\ref{fig1}(b)].  More importantly, the complexity grows rapidly around the critical points ($\mu_T=\pm 1$), changing from a convex function to a concave function at the critical points. This is further illustrated in Fig.~\ref{fig1}(c), where we plot the derivative (susceptibility) of circuit complexity with respect to $\mu_T$ .  The clear divergence at the critical points  indicates that circuit complexity is  nonanalytical at the critical points (see Supplemental Material \cite{supp} for derivation), and thus can signal the presence of a quantum phase transition. We emphasize that these features are robust signatures of phase transitions, which do not change if one chooses a different reference state in the same phase [see Figs.~\ref{fig1}(b) and (c)].

We further plot $\Delta \theta_{k_n}$ versus the momentum $k_n$, for various target states (with a fixed reference state) in Fig.~\ref{fig1}(d). When both  states are in the same phase, $\Delta \theta_{k_n}$ first increases with momentum, and finally decreases to $0$ when $k_n$ approaches $\pi$. In contrast, when $\mu_T$ is beyond its critical value, $\Delta \theta_{k_n}$ increases monotonically with momentum, and takes the maximal value of $\pi/2$ at  $k_n= \pi$.  This is closely related to the topological phase transition characterized by winding numbers, where the Bogoliubov angles  of two different states end up at the same pole (on the Bloch sphere) upon winding half of the Brillouin zone if the states belong to the same phase \cite{Alicea12}. Hence, the non-analytical nature of the circuit complexity is closely related to change of topological number (and topological phase transition).

Analytically, the derivatives of the circuit complexity \eqref{C2} can be explicitly recast into a closed contour integral over the complex variable $z=e^{i k}$ (see the Supplemental Material \cite{supp} for detailed derivations). Depending on the parameters of the target states, the poles associated with the integrand are located inside or outside the contour. When the target state goes across a phase transition, the poles sit exactly on the contour, resulting in the divergence of the derivatives of the circuit complexity at critical points \cite{supp}. Interestingly, the whole parameter space can be classified into four different phase regimes depending on which poles lie inside the contour [see Fig.~\ref{fig:branchpoints} in  Supplemental Material \cite{supp}] , which agrees exactly with the phase diagram shown in Fig.~\ref{fig1}(a).

\begin{figure}
  \centering\includegraphics[width=0.5\textwidth, height=3.8cm]{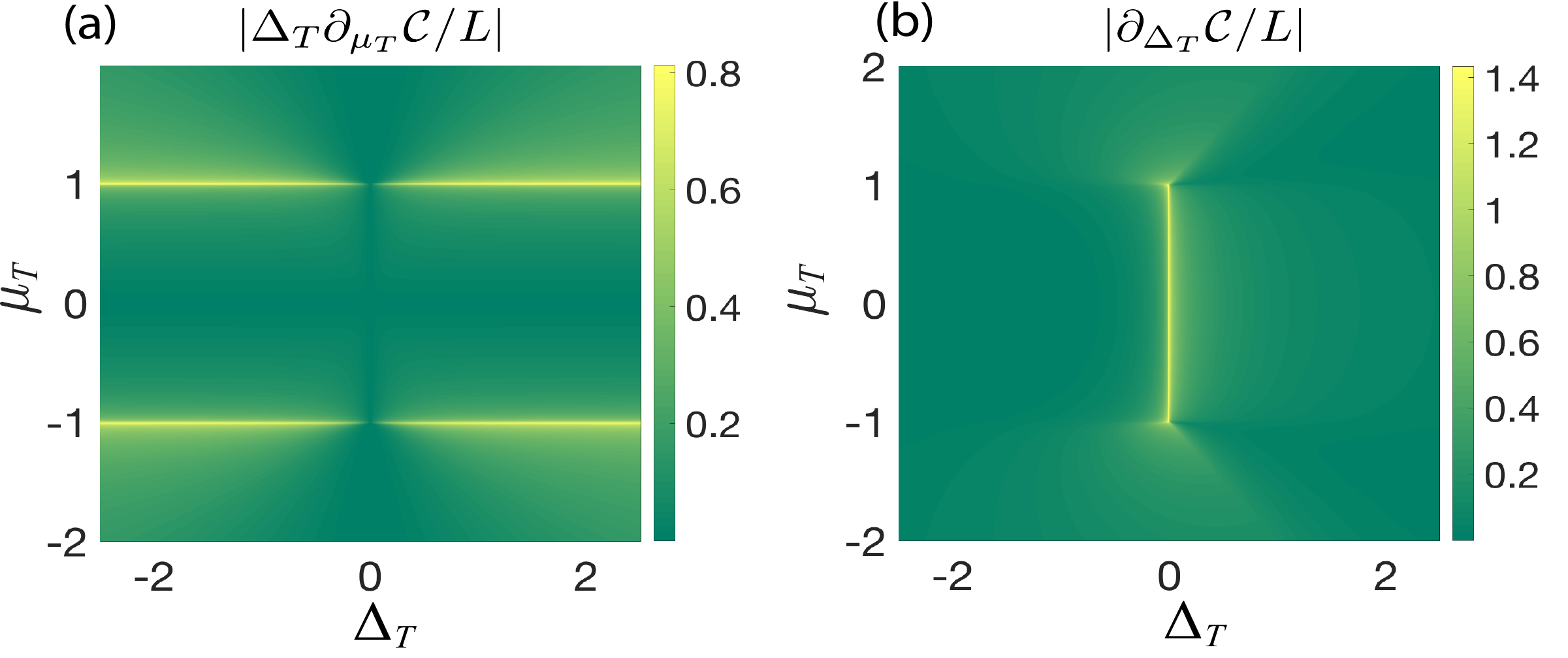}
  \caption{Derivative  of circuit complexity as a function of $\mu_T$ and $\Delta_T$.
    Panel (a) plots the derivative with respect to $\mu_T$ (in units of $1/\Delta_T$), and panel (b) plots the derivative with respect to $\Delta_T$.
    The reference state is chosen as the ground state of Eq.~\eqref{kitaev} with $\mu_R=0$ and $\Delta_R=-1$, and  $L=1000$.
  }
  
  \label{fig2}
\end{figure}

Figures~\ref{fig2}(a) and (b) show the derivative of circuit complexity with respect to $\mu_T$ and $\Delta_T$  for the whole parameter regime.  The derivatives show clear singular behavior at both the horizontal [Fig.~\ref{fig2}(a)] and vertical [Fig.~\ref{fig2}(b)] phase boundaries.   Therefore, by using the first-order derivative of complexity with respect to $\mu_T$ and $\Delta_T$, one can map out the entire equilibrium phase boundaries of the Kitaev chain. 

{\it Real-space locality of the optimal Hamiltonian.---}
Since the ground state [Eq.~(\ref{ground})] is a product of all momentum pairs, the optimal circuit connecting two different ground states corresponds to the following Hamiltonian: 
\begin{equation}
\hat{H}_c = \sum_{k>0} \hat{H}_k  = \sum_{k>0} - i \Delta \theta(k) \hat{\psi}_k^\dagger \tau_1 \hat{\psi}_k^{\phantom\dagger},
\end{equation}
where $\tau_i$ are the Pauli matrices, and $\hat{\psi}_k$ denotes the Nambu spinor
$\hat{\psi}_k = \left(\begin{array}{c}
    \hat{a}_k \\
    \hat{a}_{-k}^\dagger \\
    \end{array}\right).$ By taking a Fourier series of the above optimal Hamiltonian, one can show that the Hamiltonian can be written in real space (see Supplemental Material for details \cite{supp}): 
\begin{equation}
    \hat{H}_c = \sum_{j}\sum_{n=1}^{\infty}   \omega_n  \left( \hat{a}_j \hat{a}_{j+n}  -\textrm{H.c.} \right),
    \label{optimal_Ham}
\end{equation}
where $ {\Delta \theta(k)} = 2 \sum_{n=1}^{\infty} \omega_n \sin(nk) $. 

One crucial observation is that when the two ground states are in the same phase, $\Delta \theta(0)=\Delta \theta(\pi)=0$ [see Fig.~\ref{fig1}(d)]; hence the Fourier series of ${\Delta \theta(k)}$ converges {\it uniformly}. Therefore, the full series can be approximated by a {\it finite} order $N^{*}$ with arbitrarily small error. This immediately implies that the real-space optimal Hamiltonian~(\ref{optimal_Ham}) is local, with a {\it finite range} $N^{*}$. In sharp contrast,  if the two states belong to different phases, $\Delta \theta (\pi) = \pi/2 \neq \Delta \theta(0)=0$; the Fourier series of $\Delta \theta(k)$ converges at most pointwise. Thus the optimal Hamiltonian must be truly {\it long-range} (non-local) in real-space \cite{supp}, given that the total evolution time is chosen to be a constant [Eq.~(\ref{u_k})]. Comparing to previous works on classifying gapped phases of matter using local unitary circuits~\cite{Bravyi06, Chen10, Huang15}, our results provide an alternative approach that has a natural geometric interpretation.



{\it Complexity for dynamical topological phase transition.---}%
Dynamical  phase transitions have received tremendous interest recently \cite{Heyl17, Dalessio15, Caio15, Vajna15,Wilson16, Wang17, Caio18, Heyl18, Roy17, Paraj18, Zhang17DPT, Jurcevic17, Flasch16}. Studies on quench dynamics of circuit complexity have mostly focused on growth rates in the short-time regime \cite{Alves18, Jiang18}.  Here, we show that the long-time steady-state value of the circuit complexity following a quantum quench can be used to detect {\it dynamical} topological phase transitions. 

We take the initial state to be the ground state of a Hamiltonian $\hat{H}_i$, and consider circuit complexity growth under a sudden quench to a different Hamiltonian, $\hat{H}_f$. The reference and target states are chosen as the initial state $\ket{\Psi_i}$ and time-evolved state $\ket{\Psi(t)}$ respectively. The time-dependent $\ket{\Psi(t)}$  can be written as \cite{Quan06, Kells14}
\begin{equation}
\ket{\Psi(t)}= 
\prod_{n=0}^{\frac{L}{2}-1} \big[ \cos (\Delta \theta_{k_n})
-ie^{2i \varepsilon_{k_n}t} \sin(\Delta \theta_{k_n})\hat{A}_{k_n}^{\dagger} \hat{A}_{-k_n}^{\dagger} \big] \ket{0},
\end{equation}
where $\Delta\theta_{k_n}$ is the Bogoliubov angle difference  between eigenstates of $\hat{H}_i$ and $\hat{H}_f$, and  $\varepsilon_{k_n}$ and $\hat{A}_{k_n}$ are the energy levels  and normal mode operators, respectively,  for the post-quench Hamiltonian. Similar to the previous derivations, one can obtain the time-dependent circuit complexity,
\begin{equation}
  \label{phi_kn}
\mathcal{C}(\ket{\Psi_i} \rightarrow \ket{\Psi(t)}) =\sum_{k_n} \phi_{k_n}^2  (t)
\end{equation}
where $\phi_{k_n}  (t)= \arccos\sqrt{1-\sin^2(2\Delta \theta_{k_n}) \sin^2(\varepsilon_{k_n}t)}$.



\begin{figure}
  \centering\includegraphics[width=0.49\textwidth, height=3.6cm]{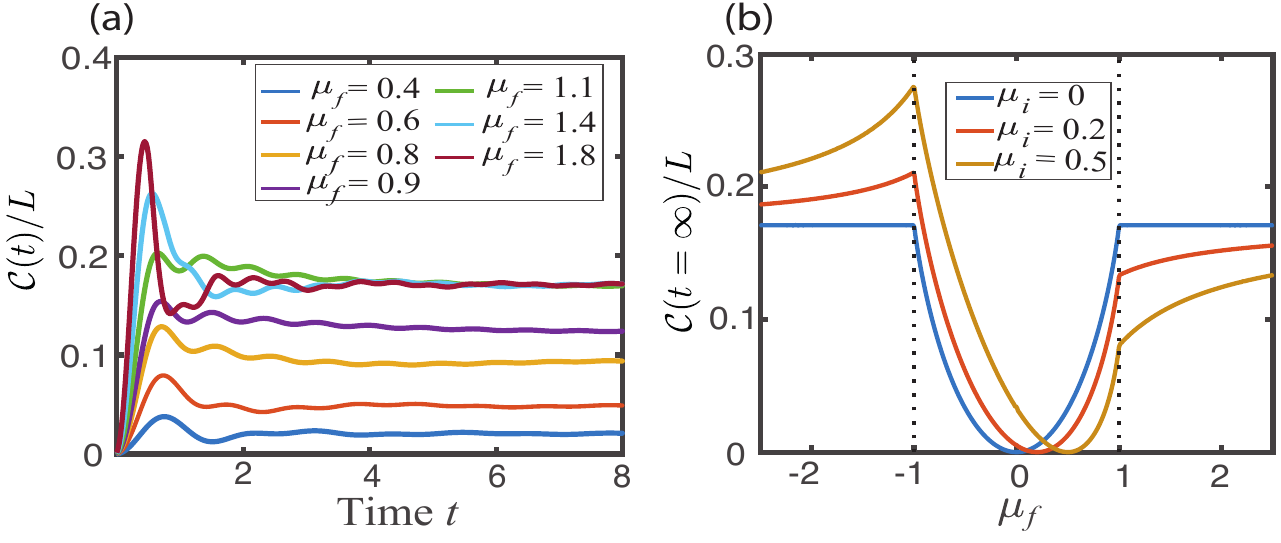}
  \caption{(a) Circuit complexity growth for various post-quench chemical potentials, $\mu_f$. The initial state (serves as the reference state) is the ground state of Eq.~\eqref{kitaev} with $\mu_i=0$.  (b) Steady-state values of complexity versus  $\mu_f$. The different  lines denote different initial/reference states.  $\Delta_i=\Delta_f=1$ and  $L=1000$ in both plots.}
  \label{fig3}
\end{figure}

 As shown in Fig.~\ref{fig3}(a), the circuit complexity first increases linearly and then oscillates \cite{Camargo18, Alves18, Jiang18} before quickly approaching a time-independent value.  The steady-state value of circuit complexity increases with $\mu_f$ of the post-quench Hamiltonian, until the phase transition occurs [Fig.~\ref{fig3}(a)]. Fig.~\ref{fig3}(b) further illustrates the long-time steady-state values of circuit complexity versus $\mu_f$ for different initial states.   The steady-state complexity clearly exhibits \textit{nonanalytical} behavior at the critical point. This behavior arises because the time-averaged value of $\phi_{k_n}(t)$ exhibits an upper bound after the phase transition (see Supplemental Material \cite{supp}),  and it is a robust feature of the dynamical phase transition regardless of the initial state.

{\it Generalization to long-range Kitaev chain and higher dimensions.---}%
We further give an example of a Kitaev chain with long-range pairing  \cite{Vodola14, Vodola16, Patrick17, Pezze17}:
\begin{equation}
\begin{split}
\hat{H}_\text{LR} =&- \frac{J}{2} \sum_{j=1}^{L}( \hat{a}_j^{\dagger}\hat{a}^{\phantom\dagger}_{j+1} +\Hc) -\mu \sum_{j=1}^{L}(\hat{a}_j^{\dagger} \hat{a}^{\phantom\dagger}_j -\frac{1}{2}) \\  &+ \frac{\Delta}{2} \sum_{j=1}^{L} \sum_{\ell=1}^{L-1}\frac{1}{d_\ell^{\alpha}}(\hat{a}_j^{\dagger} \hat{a}_{j+\ell}^{\dagger} +\Hc) , 
\end{split}
\label{kitaev_long}
\end{equation}
where $d_\ell= \min(\ell, L-\ell)$. In contrast to the short-range model, the long-range model with $\alpha < 1$ hosts topological phases with semi-integer winding numbers \cite{Vodola14, Pezze17}.
\begin{figure}
  \centering\includegraphics[width=0.49\textwidth, height=3.6cm]{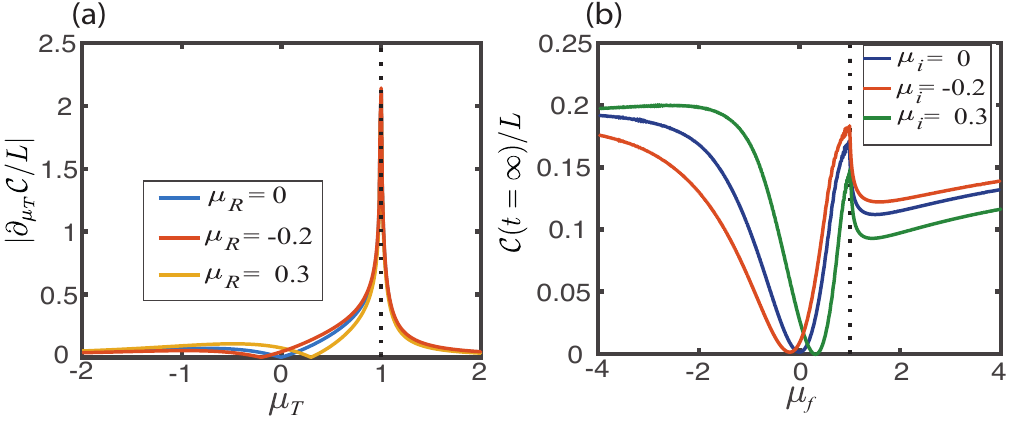}
  \caption{(a) Derivative of circuit complexity with respect to  $\mu_T$ for three different reference ground states of the long-range Kitaev chain, with $\Delta_R=\Delta_T=1.3$.  (b) Steady-state value of circuit complexity versus  $\mu_f$ for three different initial ground states, with $\Delta_i=\Delta_f=1$.  $L=1000$ and  $\alpha=0$ in both plots.
  }
  \label{fig4}
\end{figure}
As one can see, the derivative of ground state circuit complexity only   diverges at $\mu_{T}=1$ [Fig.~\ref{fig4}(a)], in contrast with Fig.~\ref{fig1}(c). This agrees perfectly with the phase diagram for the long-range interacting model, where a topological phase transition occurs only at $\mu=1$ for $\alpha=0$ \cite{ Pezze17}. Figure~\ref{fig4}(b) shows the long-time steady-state values of the circuit complexity after a sudden quench.  Again, one observes nonanalytical behavior only at $\mu_{T}=1$.


While we have so far restricted ourselves to 1D, the results we found can be readily generalized to higher dimensions \cite{Xiong19}, for example, to $p+ip$ topological superconductors in 2D. The ground state wavefunction of a $p+ip$ superconductor essentially takes the same form as Eq.~(\ref{ground}), with the momenta now being restricted to the 2D Brillouin zone, and ${\tan}(2\theta_{\bm k})=|\Delta_{\bm k}|/\varepsilon_{\bm k}$, where $\Delta_k$ and $\varepsilon_{\bm k}$ denote pairing and kinetic terms in 2D\@. The circuit complexity can still be written as $\mathcal{C}=\sum_{\bm k} |\Delta \theta_{\bm k}|^2 = \frac{L^2}{(2\pi)^2}\int {\rm d}^2{\bm k} |\Delta \theta({\bm k})|^2$. One can show again that the derivative of the circuit complexity is given by (see Supplemental Material \cite{supp})
\begin{equation}
\partial_{\mu_T}\mathcal{C}=\frac{L^2}{(2\pi)^2} \int {\rm d}^2{\bm k} \frac{\theta^T(\bm k)|\Delta({\bm k})|}{E({\bm k})^2},
\end{equation}
where $E({\bm k})^2 = \epsilon({\bm k})^2+|\Delta({\bm k})|^2$ and $\theta^T({\bm k})$ denotes the Bogoliubov angle for the target state. It is thus obvious that non-analyticity happens at the critical point where $E({\bm k})=0$ \cite{supp}.






{\it Conclusions and outlook.---}%
We use Nielsen's approach to quantify the circuit complexity of ground states and nonequilibrium steady states of the Kitaev chain with short- and long-range pairing. We find that, in both situations, circuit complexity can be used to detect topological phase transitions. The non-analytic behaviors can be generalized to higher-dimensional systems, such as $p+ip$ topological superconductors \cite{Hasan10, Qi11}.

One interesting future direction  is to use the geometric approach to quantify circuit complexity when the control Hamiltonians are constrained to be local in real-space \cite{Hyatt17, Girolami19, Huang15}, and  study its connection to quantum phase transitions \cite{Vojta03, Lieb96, Katsura62, Perk75}. It would also be of interest to investigate the circuit complexity of \emph{interacting} many-body systems.  One particular example is the XXZ spin-half chain, whose low-energy physics can be modeled by the Luttinger liquid~\cite{Haldane81, Voit95, Rahman11}.  By restricting to certain classes of gates (i.e., by imposing penalties on the cost function) \cite{Nielsen05, Jefferson17}, it might be possible to find improved methods to efficiently prepare the ground state of the XXZ model by calculating the geodesic path in gate space.

\begin{acknowledgments}
We thank Peizhi Du, Abhinav Deshpande, and Su-Kuan Chu for helpful discussions.  F.L., R.L., Z.-C.\ Y., J.R.G.\ and A.V.G.\ acknowledge support by the DoE BES QIS program (award No.\ DE-SC0019449), DoE ASCR FAR-QC (award No.\ DE-SC0020312), NSF PFCQC program, DoE ASCR Quantum Testbed Pathfinder program (award No.\ DE-SC0019040), AFOSR, NSF PFC at JQI, ARO MURI, and ARL CDQI.\@ Z.-C.\ Y.\ is also supported by AFOSR FA9550-16-1-0323, and ARO W911NF-15-1-0397.
P.T.\ and S.W.\ were supported by NIST NRC Research Postdoctoral Associateship Awards. J.B.C.\ received support from the National Science Foundation Graduate Research Fellowship Program under Grant No.\ DGE 1322106 and from the Physics Frontiers Center at JQI. This research was supported in part by the Heising-Simons Foundation, the Simons Foundation, and National Science Foundation Grant No.\ NSF PHY-1748958.
\end{acknowledgments}

\textit{Note added:}
While finalizing this manuscript, we became aware of Ref.~\cite{Ali18}, which used revivals in the circuit complexity as a qualitative probe of phase transitions in the Su-Schrieffer-Heeger model.
In contrast to that work, we have shown that the circuit complexity explicitly exhibits \emph{nonanalyticities} precisely at the critical points for the Kitaev chain. We also became aware of Ref.~\cite{Xiong19}, which numerically studied the complexity of a two- dimensional `` ${\textbf d} \cdot \tau$" model. By contrast, here we analytically study the ``$p+ip$" model, and illustrate that the closing of the gap is essential for the nonanalyticity of circuit complexity.

\bibliography{complexity}

\begin{thebibliography}{69}%
\makeatletter
\providecommand \@ifxundefined [1]{%
 \@ifx{#1\undefined}
}%
\providecommand \@ifnum [1]{%
 \ifnum #1\expandafter \@firstoftwo
 \else \expandafter \@secondoftwo
 \fi
}%
\providecommand \@ifx [1]{%
 \ifx #1\expandafter \@firstoftwo
 \else \expandafter \@secondoftwo
 \fi
}%
\providecommand \natexlab [1]{#1}%
\providecommand \enquote  [1]{``#1''}%
\providecommand \bibnamefont  [1]{#1}%
\providecommand \bibfnamefont [1]{#1}%
\providecommand \citenamefont [1]{#1}%
\providecommand \href@noop [0]{\@secondoftwo}%
\providecommand \href [0]{\begingroup \@sanitize@url \@href}%
\providecommand \@href[1]{\@@startlink{#1}\@@href}%
\providecommand \@@href[1]{\endgroup#1\@@endlink}%
\providecommand \@sanitize@url [0]{\catcode `\\12\catcode `\$12\catcode
  `\&12\catcode `\#12\catcode `\^12\catcode `\_12\catcode `\%12\relax}%
\providecommand \@@startlink[1]{}%
\providecommand \@@endlink[0]{}%
\providecommand \url  [0]{\begingroup\@sanitize@url \@url }%
\providecommand \@url [1]{\endgroup\@href {#1}{\urlprefix }}%
\providecommand \urlprefix  [0]{URL }%
\providecommand \Eprint [0]{\href }%
\providecommand \doibase [0]{http://dx.doi.org/}%
\providecommand \selectlanguage [0]{\@gobble}%
\providecommand \bibinfo  [0]{\@secondoftwo}%
\providecommand \bibfield  [0]{\@secondoftwo}%
\providecommand \translation [1]{[#1]}%
\providecommand \BibitemOpen [0]{}%
\providecommand \bibitemStop [0]{}%
\providecommand \bibitemNoStop [0]{.\EOS\space}%
\providecommand \EOS [0]{\spacefactor3000\relax}%
\providecommand \BibitemShut  [1]{\csname bibitem#1\endcsname}%
\let\auto@bib@innerbib\@empty
\bibitem [{\citenamefont {Watrous}(2009)}]{Watrous2009}%
  \BibitemOpen
  \bibfield  {author} {\bibinfo {author} {\bibfnamefont {J.}~\bibnamefont
  {Watrous}},\ }\enquote {\bibinfo {title} {Quantum computational
  complexity},}\ in\ \href {\doibase 10.1007/978-0-387-30440-3_428} {\emph
  {\bibinfo {booktitle} {Encyclopedia of Complexity and Systems Science}}},\
  \bibinfo {editor} {edited by\ \bibinfo {editor} {\bibfnamefont {Robert~A.}\
  \bibnamefont {Meyers}}}\ (\bibinfo  {publisher} {Springer New York},\
  \bibinfo {address} {New York, NY},\ \bibinfo {year} {2009})\ pp.\ \bibinfo
  {pages} {7174--7201}\BibitemShut {NoStop}%
\bibitem [{\citenamefont {{Aaronson}}()}]{Aaronson16}%
  \BibitemOpen
  \bibfield  {author} {\bibinfo {author} {\bibfnamefont {S.}~\bibnamefont
  {{Aaronson}}},\ }\bibfield  {title} {\enquote {\bibinfo {title} {{The
  Complexity of Quantum States and Transformations: From Quantum Money to Black
  Holes}},}\ }\href@noop {} {\ }\Eprint {http://arxiv.org/abs/1607.05256}
  {arXiv:1607.05256} \BibitemShut {NoStop}%
\bibitem [{\citenamefont {{Nielsen}}()}]{Nielsen05}%
  \BibitemOpen
  \bibfield  {author} {\bibinfo {author} {\bibfnamefont {M.~A.}\ \bibnamefont
  {{Nielsen}}},\ }\bibfield  {title} {\enquote {\bibinfo {title} {{A geometric
  approach to quantum circuit lower bounds}},}\ }\href@noop {} {\ }\Eprint
  {http://arxiv.org/abs/quant-ph/0502070} {arXiv:quant-ph/0502070} \BibitemShut
  {NoStop}%
\bibitem [{\citenamefont {{Nielsen}}\ \emph {et~al.}(2006)\citenamefont
  {{Nielsen}}, \citenamefont {{Dowling}}, \citenamefont {{Gu}},\ and\
  \citenamefont {{Doherty}}}]{Nielsen06}%
  \BibitemOpen
  \bibfield  {author} {\bibinfo {author} {\bibfnamefont {M.~A.}\ \bibnamefont
  {{Nielsen}}}, \bibinfo {author} {\bibfnamefont {M.~R.}\ \bibnamefont
  {{Dowling}}}, \bibinfo {author} {\bibfnamefont {M.}~\bibnamefont {{Gu}}}, \
  and\ \bibinfo {author} {\bibfnamefont {A.~C.}\ \bibnamefont {{Doherty}}},\
  }\bibfield  {title} {\enquote {\bibinfo {title} {{Quantum Computation as
  Geometry}},}\ }\href {\doibase 10.1126/science.1121541} {\bibfield  {journal}
  {\bibinfo  {journal} {Science}\ }\textbf {\bibinfo {volume} {311}},\ \bibinfo
  {pages} {1133} (\bibinfo {year} {2006})}\BibitemShut {NoStop}%
\bibitem [{\citenamefont {{Dowling}}\ and\ \citenamefont
  {{Nielsen}}()}]{Dowling07}%
  \BibitemOpen
  \bibfield  {author} {\bibinfo {author} {\bibfnamefont {M.~R.}\ \bibnamefont
  {{Dowling}}}\ and\ \bibinfo {author} {\bibfnamefont {M.~A.}\ \bibnamefont
  {{Nielsen}}},\ }\bibfield  {title} {\enquote {\bibinfo {title} {{The geometry
  of quantum computation}},}\ }\href@noop {} {\ }\Eprint
  {http://arxiv.org/abs/quant-ph/0701004} {arXiv:quant-ph/0701004} \BibitemShut
  {NoStop}%
\bibitem [{\citenamefont {{Jefferson}}\ and\ \citenamefont
  {{Myers}}(2017)}]{Jefferson17}%
  \BibitemOpen
  \bibfield  {author} {\bibinfo {author} {\bibfnamefont {R.~A.}\ \bibnamefont
  {{Jefferson}}}\ and\ \bibinfo {author} {\bibfnamefont {R.~C.}\ \bibnamefont
  {{Myers}}},\ }\bibfield  {title} {\enquote {\bibinfo {title} {{Circuit
  complexity in quantum field theory}},}\ }\href {\doibase
  10.1007/JHEP10(2017)107} {\bibfield  {journal} {\bibinfo  {journal} {J. High
  Energy Phys.}\ }\textbf {\bibinfo {volume} {10}},\ \bibinfo {eid} {107}
  (\bibinfo {year} {2017})}\BibitemShut {NoStop}%
\bibitem [{\citenamefont {Yang}(2018)}]{Yang18}%
  \BibitemOpen
  \bibfield  {author} {\bibinfo {author} {\bibfnamefont {R.-Q.}\ \bibnamefont
  {Yang}},\ }\bibfield  {title} {\enquote {\bibinfo {title} {Complexity for
  quantum field theory states and applications to thermofield double states},}\
  }\href {\doibase 10.1103/PhysRevD.97.066004} {\bibfield  {journal} {\bibinfo
  {journal} {Phys. Rev. D}\ }\textbf {\bibinfo {volume} {97}},\ \bibinfo
  {pages} {066004} (\bibinfo {year} {2018})}\BibitemShut {NoStop}%
\bibitem [{\citenamefont {{Guo}}\ \emph {et~al.}(2018)\citenamefont {{Guo}},
  \citenamefont {{Hernandez}}, \citenamefont {{Myers}},\ and\ \citenamefont
  {{Ruan}}}]{Guo18}%
  \BibitemOpen
  \bibfield  {author} {\bibinfo {author} {\bibfnamefont {M.}~\bibnamefont
  {{Guo}}}, \bibinfo {author} {\bibfnamefont {J.}~\bibnamefont {{Hernandez}}},
  \bibinfo {author} {\bibfnamefont {R.~C.}\ \bibnamefont {{Myers}}}, \ and\
  \bibinfo {author} {\bibfnamefont {S.-M.}\ \bibnamefont {{Ruan}}},\ }\bibfield
   {title} {\enquote {\bibinfo {title} {{Circuit complexity for coherent
  states}},}\ }\href {\doibase 10.1007/JHEP10(2018)011} {\bibfield  {journal}
  {\bibinfo  {journal} {J. High Energy Phys.}\ }\textbf {\bibinfo {volume}
  {10}},\ \bibinfo {eid} {11} (\bibinfo {year} {2018})}\BibitemShut {NoStop}%
\bibitem [{\citenamefont {{Camargo}}\ \emph {et~al.}()\citenamefont
  {{Camargo}}, \citenamefont {{Caputa}}, \citenamefont {{Das}}, \citenamefont
  {{Heller}},\ and\ \citenamefont {{Jefferson}}}]{Camargo18}%
  \BibitemOpen
  \bibfield  {author} {\bibinfo {author} {\bibfnamefont {H.~A.}\ \bibnamefont
  {{Camargo}}}, \bibinfo {author} {\bibfnamefont {P.}~\bibnamefont {{Caputa}}},
  \bibinfo {author} {\bibfnamefont {D.}~\bibnamefont {{Das}}}, \bibinfo
  {author} {\bibfnamefont {M.~P.}\ \bibnamefont {{Heller}}}, \ and\ \bibinfo
  {author} {\bibfnamefont {R.}~\bibnamefont {{Jefferson}}},\ }\bibfield
  {title} {\enquote {\bibinfo {title} {{Complexity as a novel probe of quantum
  quenches: universal scalings and purifications}},}\ }\href@noop {} {\
  }\Eprint {http://arxiv.org/abs/1807.07075} {arXiv:1807.07075} \BibitemShut
  {NoStop}%
\bibitem [{\citenamefont {{Alves}}\ and\ \citenamefont
  {{Camilo}}(2018)}]{Alves18}%
  \BibitemOpen
  \bibfield  {author} {\bibinfo {author} {\bibfnamefont {D.~W.~F.}\
  \bibnamefont {{Alves}}}\ and\ \bibinfo {author} {\bibfnamefont
  {G.}~\bibnamefont {{Camilo}}},\ }\bibfield  {title} {\enquote {\bibinfo
  {title} {{Evolution of complexity following a quantum quench in free field
  theory}},}\ }\href {\doibase 10.1007/JHEP06(2018)029} {\bibfield  {journal}
  {\bibinfo  {journal} {J. High Energy Phys.}\ }\textbf {\bibinfo {volume}
  {6}},\ \bibinfo {eid} {29} (\bibinfo {year} {2018})}\BibitemShut {NoStop}%
\bibitem [{\citenamefont {{Chapman}}\ \emph {et~al.}()\citenamefont
  {{Chapman}}, \citenamefont {{Eisert}}, \citenamefont {{Hackl}}, \citenamefont
  {{Heller}}, \citenamefont {{Jefferson}}, \citenamefont {{Marrochio}},\ and\
  \citenamefont {{Myers}}}]{Chapman18}%
  \BibitemOpen
  \bibfield  {author} {\bibinfo {author} {\bibfnamefont {S.}~\bibnamefont
  {{Chapman}}}, \bibinfo {author} {\bibfnamefont {J.}~\bibnamefont {{Eisert}}},
  \bibinfo {author} {\bibfnamefont {L.}~\bibnamefont {{Hackl}}}, \bibinfo
  {author} {\bibfnamefont {M.~P.}\ \bibnamefont {{Heller}}}, \bibinfo {author}
  {\bibfnamefont {R.}~\bibnamefont {{Jefferson}}}, \bibinfo {author}
  {\bibfnamefont {H.}~\bibnamefont {{Marrochio}}}, \ and\ \bibinfo {author}
  {\bibfnamefont {R.~C.}\ \bibnamefont {{Myers}}},\ }\bibfield  {title}
  {\enquote {\bibinfo {title} {{Complexity and entanglement for thermofield
  double states}},}\ }\href@noop {} {\ }\Eprint
  {http://arxiv.org/abs/1810.05151} {arXiv:1810.05151} \BibitemShut {NoStop}%
\bibitem [{\citenamefont {{Hackl}}\ and\ \citenamefont
  {{Myers}}(2018)}]{Hackl18}%
  \BibitemOpen
  \bibfield  {author} {\bibinfo {author} {\bibfnamefont {L.}~\bibnamefont
  {{Hackl}}}\ and\ \bibinfo {author} {\bibfnamefont {R.~C.}\ \bibnamefont
  {{Myers}}},\ }\bibfield  {title} {\enquote {\bibinfo {title} {{Circuit
  complexity for free fermions}},}\ }\href {\doibase 10.1007/JHEP07(2018)139}
  {\bibfield  {journal} {\bibinfo  {journal} {J. High Energy Phys.}\ }\textbf
  {\bibinfo {volume} {7}},\ \bibinfo {eid} {139} (\bibinfo {year}
  {2018})}\BibitemShut {NoStop}%
\bibitem [{\citenamefont {{Khan}}\ \emph {et~al.}()\citenamefont {{Khan}},
  \citenamefont {{Krishnan}},\ and\ \citenamefont {{Sharma}}}]{Khan18}%
  \BibitemOpen
  \bibfield  {author} {\bibinfo {author} {\bibfnamefont {R.}~\bibnamefont
  {{Khan}}}, \bibinfo {author} {\bibfnamefont {C.}~\bibnamefont {{Krishnan}}},
  \ and\ \bibinfo {author} {\bibfnamefont {S.}~\bibnamefont {{Sharma}}},\
  }\bibfield  {title} {\enquote {\bibinfo {title} {{Circuit Complexity in
  Fermionic Field Theory}},}\ }\href@noop {} {\ }\Eprint
  {http://arxiv.org/abs/1801.07620} {arXiv:1801.07620} \BibitemShut {NoStop}%
\bibitem [{\citenamefont {{Reynolds}}\ and\ \citenamefont
  {{Ross}}(2018)}]{Rey17}%
  \BibitemOpen
  \bibfield  {author} {\bibinfo {author} {\bibfnamefont {A.~P.}\ \bibnamefont
  {{Reynolds}}}\ and\ \bibinfo {author} {\bibfnamefont {S.~F.}\ \bibnamefont
  {{Ross}}},\ }\bibfield  {title} {\enquote {\bibinfo {title} {{Complexity of
  the AdS soliton}},}\ }\href {\doibase 10.1088/1361-6382/aab32d} {\bibfield
  {journal} {\bibinfo  {journal} {Class. Quantum Grav}\ }\textbf {\bibinfo
  {volume} {35}},\ \bibinfo {eid} {095006} (\bibinfo {year}
  {2018})}\BibitemShut {NoStop}%
\bibitem [{\citenamefont {{Jiang}}\ \emph {et~al.}()\citenamefont {{Jiang}},
  \citenamefont {{Shan}},\ and\ \citenamefont {{Yang}}}]{Jiang18}%
  \BibitemOpen
  \bibfield  {author} {\bibinfo {author} {\bibfnamefont {J.}~\bibnamefont
  {{Jiang}}}, \bibinfo {author} {\bibfnamefont {J.}~\bibnamefont {{Shan}}}, \
  and\ \bibinfo {author} {\bibfnamefont {J.}~\bibnamefont {{Yang}}},\
  }\bibfield  {title} {\enquote {\bibinfo {title} {{Circuit complexity for free
  Fermion with a mass quench}},}\ }\href@noop {} {\ }\Eprint
  {http://arxiv.org/abs/1810.00537} {arXiv:1810.00537} \BibitemShut {NoStop}%
\bibitem [{\citenamefont {{Yang}}\ \emph
  {et~al.}(2019{\natexlab{a}})\citenamefont {{Yang}}, \citenamefont {{An}},
  \citenamefont {{Niu}}, \citenamefont {{Zhang}},\ and\ \citenamefont
  {{Kim}}}]{YRQ19_1}%
  \BibitemOpen
  \bibfield  {author} {\bibinfo {author} {\bibfnamefont {R.~Q.}\ \bibnamefont
  {{Yang}}}, \bibinfo {author} {\bibfnamefont {Y.~S.}\ \bibnamefont {{An}}},
  \bibinfo {author} {\bibfnamefont {C.}~\bibnamefont {{Niu}}}, \bibinfo
  {author} {\bibfnamefont {C.~Y.}\ \bibnamefont {{Zhang}}}, \ and\ \bibinfo
  {author} {\bibfnamefont {K.~Y.}\ \bibnamefont {{Kim}}},\ }\bibfield  {title}
  {\enquote {\bibinfo {title} {{Principles and symmetries of complexity in
  quantum field theory}},}\ }\href {\doibase 10.1140/epjc/s10052-019-6600-3}
  {\bibfield  {journal} {\bibinfo  {journal} {Eur. Phys. J. C}\ }\textbf
  {\bibinfo {volume} {79}},\ \bibinfo {eid} {109} (\bibinfo {year}
  {2019}{\natexlab{a}})}\BibitemShut {NoStop}%
\bibitem [{\citenamefont {{Yang}}\ \emph
  {et~al.}(2019{\natexlab{b}})\citenamefont {{Yang}}, \citenamefont {{An}},
  \citenamefont {{Niu}}, \citenamefont {{Zhang}},\ and\ \citenamefont
  {{Kim}}}]{YRQ19_2}%
  \BibitemOpen
  \bibfield  {author} {\bibinfo {author} {\bibfnamefont {R.~Q.}\ \bibnamefont
  {{Yang}}}, \bibinfo {author} {\bibfnamefont {Y.~S.}\ \bibnamefont {{An}}},
  \bibinfo {author} {\bibfnamefont {C.}~\bibnamefont {{Niu}}}, \bibinfo
  {author} {\bibfnamefont {C.~Y.}\ \bibnamefont {{Zhang}}}, \ and\ \bibinfo
  {author} {\bibfnamefont {K.~Y.}\ \bibnamefont {{Kim}}},\ }\bibfield  {title}
  {\enquote {\bibinfo {title} {{More on complexity of operators in quantum
  field theory}},}\ }\href {\doibase 10.1007/JHEP03(2019)161} {\bibfield
  {journal} {\bibinfo  {journal} {J. High Energy Phys.}\ }\textbf {\bibinfo
  {volume} {2019}},\ \bibinfo {eid} {161} (\bibinfo {year}
  {2019}{\natexlab{b}})}\BibitemShut {NoStop}%
\bibitem [{\citenamefont {{Yang}}\ and\ \citenamefont {{Kim}}(2019)}]{YRQ19_3}%
  \BibitemOpen
  \bibfield  {author} {\bibinfo {author} {\bibfnamefont {R.~Q.}\ \bibnamefont
  {{Yang}}}\ and\ \bibinfo {author} {\bibfnamefont {K.~Y.}\ \bibnamefont
  {{Kim}}},\ }\bibfield  {title} {\enquote {\bibinfo {title} {{Complexity of
  operators generated by quantum mechanical Hamiltonians}},}\ }\href {\doibase
  10.1007/JHEP03(2019)010} {\bibfield  {journal} {\bibinfo  {journal} {J. High
  Energy Phys.}\ }\textbf {\bibinfo {volume} {2019}},\ \bibinfo {eid} {10}
  (\bibinfo {year} {2019})}\BibitemShut {NoStop}%
\bibitem [{\citenamefont {{Stanford}}\ and\ \citenamefont
  {{Susskind}}(2014)}]{Stanford14}%
  \BibitemOpen
  \bibfield  {author} {\bibinfo {author} {\bibfnamefont {D.}~\bibnamefont
  {{Stanford}}}\ and\ \bibinfo {author} {\bibfnamefont {L.}~\bibnamefont
  {{Susskind}}},\ }\bibfield  {title} {\enquote {\bibinfo {title} {{Complexity
  and shock wave geometries}},}\ }\href {\doibase 10.1103/PhysRevD.90.126007}
  {\bibfield  {journal} {\bibinfo  {journal} {\prd}\ }\textbf {\bibinfo
  {volume} {90}},\ \bibinfo {eid} {126007} (\bibinfo {year}
  {2014})}\BibitemShut {NoStop}%
\bibitem [{\citenamefont {{Susskind}}()}]{Susskind14}%
  \BibitemOpen
  \bibfield  {author} {\bibinfo {author} {\bibfnamefont {L.}~\bibnamefont
  {{Susskind}}},\ }\bibfield  {title} {\enquote {\bibinfo {title}
  {{Computational Complexity and Black Hole Horizons}},}\ }\href@noop {} {\
  }\Eprint {http://arxiv.org/abs/1402.5674} {arXiv:1402.5674} \BibitemShut
  {NoStop}%
\bibitem [{\citenamefont {{Brown}}\ \emph {et~al.}()\citenamefont {{Brown}},
  \citenamefont {{Roberts}}, \citenamefont {{Susskind}}, \citenamefont
  {{Swingle}},\ and\ \citenamefont {{Zhao}}}]{Brown15}%
  \BibitemOpen
  \bibfield  {author} {\bibinfo {author} {\bibfnamefont {A.~R.}\ \bibnamefont
  {{Brown}}}, \bibinfo {author} {\bibfnamefont {D.~A.}\ \bibnamefont
  {{Roberts}}}, \bibinfo {author} {\bibfnamefont {L.}~\bibnamefont
  {{Susskind}}}, \bibinfo {author} {\bibfnamefont {B.}~\bibnamefont
  {{Swingle}}}, \ and\ \bibinfo {author} {\bibfnamefont {Y.}~\bibnamefont
  {{Zhao}}},\ }\bibfield  {title} {\enquote {\bibinfo {title} {{Complexity
  Equals Action}},}\ }\href@noop {} {\ }\Eprint
  {http://arxiv.org/abs/1509.07876} {arXiv:1509.07876} \BibitemShut {NoStop}%
\bibitem [{\citenamefont {{Brown}}\ \emph {et~al.}(2016)\citenamefont
  {{Brown}}, \citenamefont {{Roberts}}, \citenamefont {{Susskind}},
  \citenamefont {{Swingle}},\ and\ \citenamefont {{Zhao}}}]{Brown16}%
  \BibitemOpen
  \bibfield  {author} {\bibinfo {author} {\bibfnamefont {A.~R.}\ \bibnamefont
  {{Brown}}}, \bibinfo {author} {\bibfnamefont {D.~A.}\ \bibnamefont
  {{Roberts}}}, \bibinfo {author} {\bibfnamefont {L.}~\bibnamefont
  {{Susskind}}}, \bibinfo {author} {\bibfnamefont {B.}~\bibnamefont
  {{Swingle}}}, \ and\ \bibinfo {author} {\bibfnamefont {Y.}~\bibnamefont
  {{Zhao}}},\ }\bibfield  {title} {\enquote {\bibinfo {title} {{Complexity,
  action, and black holes}},}\ }\href {\doibase 10.1103/PhysRevD.93.086006}
  {\bibfield  {journal} {\bibinfo  {journal} {\prd}\ }\textbf {\bibinfo
  {volume} {93}},\ \bibinfo {eid} {086006} (\bibinfo {year}
  {2016})}\BibitemShut {NoStop}%
\bibitem [{\citenamefont {{Chapman}}\ \emph {et~al.}(2018)\citenamefont
  {{Chapman}}, \citenamefont {{Heller}}, \citenamefont {{Marrochio}},\ and\
  \citenamefont {{Pastawski}}}]{Chapman18PRL}%
  \BibitemOpen
  \bibfield  {author} {\bibinfo {author} {\bibfnamefont {S.}~\bibnamefont
  {{Chapman}}}, \bibinfo {author} {\bibfnamefont {M.~P.}\ \bibnamefont
  {{Heller}}}, \bibinfo {author} {\bibfnamefont {H.}~\bibnamefont
  {{Marrochio}}}, \ and\ \bibinfo {author} {\bibfnamefont {F.}~\bibnamefont
  {{Pastawski}}},\ }\bibfield  {title} {\enquote {\bibinfo {title} {{Toward a
  Definition of Complexity for Quantum Field Theory States}},}\ }\href
  {\doibase 10.1103/PhysRevLett.120.121602} {\bibfield  {journal} {\bibinfo
  {journal} {Phys. Rev. Lett}\ }\textbf {\bibinfo {volume} {120}},\ \bibinfo
  {eid} {121602} (\bibinfo {year} {2018})}\BibitemShut {NoStop}%
\bibitem [{\citenamefont {{Caputa}}\ and\ \citenamefont
  {{Magan}}()}]{Caputa18}%
  \BibitemOpen
  \bibfield  {author} {\bibinfo {author} {\bibfnamefont {P.}~\bibnamefont
  {{Caputa}}}\ and\ \bibinfo {author} {\bibfnamefont {J.~M.}\ \bibnamefont
  {{Magan}}},\ }\bibfield  {title} {\enquote {\bibinfo {title} {{Quantum
  Computation as Gravity}},}\ }\href@noop {} {\ }\Eprint
  {http://arxiv.org/abs/1807.04422} {arXiv:1807.04422} \BibitemShut {NoStop}%
\bibitem [{\citenamefont {{Vojta}}(2003)}]{Vojta03}%
  \BibitemOpen
  \bibfield  {author} {\bibinfo {author} {\bibfnamefont {M.}~\bibnamefont
  {{Vojta}}},\ }\bibfield  {title} {\enquote {\bibinfo {title} {{Quantum phase
  transitions}},}\ }\href {\doibase 10.1088/0034-4885/66/12/R01} {\bibfield
  {journal} {\bibinfo  {journal} {Rep. Prog. Phys}\ }\textbf {\bibinfo {volume}
  {66}},\ \bibinfo {pages} {2069} (\bibinfo {year} {2003})}\BibitemShut
  {NoStop}%
\bibitem [{\citenamefont {Caneva}\ \emph {et~al.}(2007)\citenamefont {Caneva},
  \citenamefont {Fazio},\ and\ \citenamefont {Santoro}}]{Caneva}%
  \BibitemOpen
  \bibfield  {author} {\bibinfo {author} {\bibfnamefont {T.}~\bibnamefont
  {Caneva}}, \bibinfo {author} {\bibfnamefont {R.}~\bibnamefont {Fazio}}, \
  and\ \bibinfo {author} {\bibfnamefont {G.~E.}\ \bibnamefont {Santoro}},\
  }\bibfield  {title} {\enquote {\bibinfo {title} {{Adiabatic quantum dynamics
  of a random Ising chain across its quantum critical point}},}\ }\href
  {\doibase 10.1103/PhysRevB.76.144427} {\bibfield  {journal} {\bibinfo
  {journal} {Phys. Rev. B}\ }\textbf {\bibinfo {volume} {76}},\ \bibinfo
  {pages} {144427} (\bibinfo {year} {2007})}\BibitemShut {NoStop}%
\bibitem [{\citenamefont {{S{\o}rensen}}\ \emph {et~al.}(2010)\citenamefont
  {{S{\o}rensen}}, \citenamefont {{Altman}}, \citenamefont {{Gullans}},
  \citenamefont {{Porto}}, \citenamefont {{Lukin}},\ and\ \citenamefont
  {{Demler}}}]{Sorensen10}%
  \BibitemOpen
  \bibfield  {author} {\bibinfo {author} {\bibfnamefont {A.~S.}\ \bibnamefont
  {{S{\o}rensen}}}, \bibinfo {author} {\bibfnamefont {E.}~\bibnamefont
  {{Altman}}}, \bibinfo {author} {\bibfnamefont {M.}~\bibnamefont {{Gullans}}},
  \bibinfo {author} {\bibfnamefont {J.~V.}\ \bibnamefont {{Porto}}}, \bibinfo
  {author} {\bibfnamefont {M.~D.}\ \bibnamefont {{Lukin}}}, \ and\ \bibinfo
  {author} {\bibfnamefont {E.}~\bibnamefont {{Demler}}},\ }\bibfield  {title}
  {\enquote {\bibinfo {title} {{Adiabatic preparation of many-body states in
  optical lattices}},}\ }\href {\doibase 10.1103/PhysRevA.81.061603} {\bibfield
   {journal} {\bibinfo  {journal} {\pra}\ }\textbf {\bibinfo {volume} {81}},\
  \bibinfo {eid} {061603} (\bibinfo {year} {2010})}\BibitemShut {NoStop}%
\bibitem [{\citenamefont {Kitaev}(2001)}]{Kitaev01}%
  \BibitemOpen
  \bibfield  {author} {\bibinfo {author} {\bibfnamefont {A.~Y.}\ \bibnamefont
  {Kitaev}},\ }\bibfield  {title} {\enquote {\bibinfo {title} {{Unpaired
  Majorana fermions in quantum wires}},}\ }\href {\doibase
  10.1070/1063-7869/44/10S/S29} {\bibfield  {journal} {\bibinfo  {journal}
  {Phys.-Uspekhi}\ }\textbf {\bibinfo {volume} {44}},\ \bibinfo {pages} {131}
  (\bibinfo {year} {2001})}\BibitemShut {NoStop}%
\bibitem [{\citenamefont {{Alicea}}(2012)}]{Alicea12}%
  \BibitemOpen
  \bibfield  {author} {\bibinfo {author} {\bibfnamefont {J.}~\bibnamefont
  {{Alicea}}},\ }\bibfield  {title} {\enquote {\bibinfo {title} {{New
  directions in the pursuit of Majorana fermions in solid state systems}},}\
  }\href {\doibase 10.1088/0034-4885/75/7/076501} {\bibfield  {journal}
  {\bibinfo  {journal} {Rep. Prog. Phys}\ }\textbf {\bibinfo {volume} {75}},\
  \bibinfo {eid} {076501} (\bibinfo {year} {2012})}\BibitemShut {NoStop}%
\bibitem [{\citenamefont {{Alicea}}\ \emph {et~al.}(2011)\citenamefont
  {{Alicea}}, \citenamefont {{Oreg}}, \citenamefont {{Refael}}, \citenamefont
  {{von Oppen}},\ and\ \citenamefont {{Fisher}}}]{Jason11}%
  \BibitemOpen
  \bibfield  {author} {\bibinfo {author} {\bibfnamefont {J.}~\bibnamefont
  {{Alicea}}}, \bibinfo {author} {\bibfnamefont {Y.}~\bibnamefont {{Oreg}}},
  \bibinfo {author} {\bibfnamefont {G.}~\bibnamefont {{Refael}}}, \bibinfo
  {author} {\bibfnamefont {F.}~\bibnamefont {{von Oppen}}}, \ and\ \bibinfo
  {author} {\bibfnamefont {M.~P.~A.}\ \bibnamefont {{Fisher}}},\ }\bibfield
  {title} {\enquote {\bibinfo {title} {{Non-Abelian statistics and topological
  quantum information processing in 1D wire networks}},}\ }\href {\doibase
  10.1038/nphys1915} {\bibfield  {journal} {\bibinfo  {journal} {Nat. Phys}\
  }\textbf {\bibinfo {volume} {7}},\ \bibinfo {pages} {412} (\bibinfo {year}
  {2011})}\BibitemShut {NoStop}%
\bibitem [{\citenamefont {Sau}\ \emph {et~al.}(2010)\citenamefont {Sau},
  \citenamefont {Lutchyn}, \citenamefont {Tewari},\ and\ \citenamefont
  {Das~Sarma}}]{Sau10}%
  \BibitemOpen
  \bibfield  {author} {\bibinfo {author} {\bibfnamefont {J.~D.}\ \bibnamefont
  {Sau}}, \bibinfo {author} {\bibfnamefont {R.~M.}\ \bibnamefont {Lutchyn}},
  \bibinfo {author} {\bibfnamefont {S.}~\bibnamefont {Tewari}}, \ and\ \bibinfo
  {author} {\bibfnamefont {S.}~\bibnamefont {Das~Sarma}},\ }\bibfield  {title}
  {\enquote {\bibinfo {title} {Generic new platform for topological quantum
  computation using semiconductor heterostructures},}\ }\href {\doibase
  10.1103/PhysRevLett.104.040502} {\bibfield  {journal} {\bibinfo  {journal}
  {Phys. Rev. Lett.}\ }\textbf {\bibinfo {volume} {104}},\ \bibinfo {pages}
  {040502} (\bibinfo {year} {2010})}\BibitemShut {NoStop}%
\bibitem [{\citenamefont {Oreg}\ \emph {et~al.}(2010)\citenamefont {Oreg},
  \citenamefont {Refael},\ and\ \citenamefont {von Oppen}}]{Oreg10}%
  \BibitemOpen
  \bibfield  {author} {\bibinfo {author} {\bibfnamefont {Y.}~\bibnamefont
  {Oreg}}, \bibinfo {author} {\bibfnamefont {G.}~\bibnamefont {Refael}}, \ and\
  \bibinfo {author} {\bibfnamefont {F.}~\bibnamefont {von Oppen}},\ }\bibfield
  {title} {\enquote {\bibinfo {title} {{Helical Liquids and Majorana Bound
  States in Quantum Wires}},}\ }\href {\doibase 10.1103/PhysRevLett.105.177002}
  {\bibfield  {journal} {\bibinfo  {journal} {Phys. Rev. Lett.}\ }\textbf
  {\bibinfo {volume} {105}},\ \bibinfo {pages} {177002} (\bibinfo {year}
  {2010})}\BibitemShut {NoStop}%
\bibitem [{\citenamefont {Lutchyn}\ \emph {et~al.}(2010)\citenamefont
  {Lutchyn}, \citenamefont {Sau},\ and\ \citenamefont {Das~Sarma}}]{Lutch10}%
  \BibitemOpen
  \bibfield  {author} {\bibinfo {author} {\bibfnamefont {R.~M.}\ \bibnamefont
  {Lutchyn}}, \bibinfo {author} {\bibfnamefont {J.~D.}\ \bibnamefont {Sau}}, \
  and\ \bibinfo {author} {\bibfnamefont {S.}~\bibnamefont {Das~Sarma}},\
  }\bibfield  {title} {\enquote {\bibinfo {title} {Majorana fermions and a
  topological phase transition in semiconductor-superconductor
  heterostructures},}\ }\href {\doibase 10.1103/PhysRevLett.105.077001}
  {\bibfield  {journal} {\bibinfo  {journal} {Phys. Rev. Lett.}\ }\textbf
  {\bibinfo {volume} {105}},\ \bibinfo {pages} {077001} (\bibinfo {year}
  {2010})}\BibitemShut {NoStop}%
\bibitem [{sup()}]{supp}%
  \BibitemOpen
  \href@noop {} {}\bibinfo {note} {See Supplemental Material for derivations of
  circuit complexity for a pair of fermions, derivations of the nonanalyticity
  of circuit complexity at critical points, details of real-space behavior of
  optimal control Hamiltonians, numerics for the nonanalyticity of circuit
  complexity after quantum quenches, and circuit complexity for 2D $p+ip$
  topological superconductors.}\BibitemShut {Stop}%
\bibitem [{Note1()}]{Note1}%
  \BibitemOpen
  \bibinfo {note} {In such a space, each state is represented by one point,
  with its coordinates labeled by the Bogoliubov angles, i.e.\ $(\theta _{k_0},
  \theta _{k_1}, \protect \dots , \theta _{k_{L/2-1}})$}\BibitemShut {NoStop}%
\bibitem [{\citenamefont {Bravyi}\ \emph {et~al.}(2006)\citenamefont {Bravyi},
  \citenamefont {Hastings},\ and\ \citenamefont {Verstraete}}]{Bravyi06}%
  \BibitemOpen
  \bibfield  {author} {\bibinfo {author} {\bibfnamefont {S.}~\bibnamefont
  {Bravyi}}, \bibinfo {author} {\bibfnamefont {M.~B.}\ \bibnamefont
  {Hastings}}, \ and\ \bibinfo {author} {\bibfnamefont {F.}~\bibnamefont
  {Verstraete}},\ }\bibfield  {title} {\enquote {\bibinfo {title}
  {Lieb-robinson bounds and the generation of correlations and topological
  quantum order},}\ }\href {\doibase 10.1103/PhysRevLett.97.050401} {\bibfield
  {journal} {\bibinfo  {journal} {Phys. Rev. Lett.}\ }\textbf {\bibinfo
  {volume} {97}},\ \bibinfo {pages} {050401} (\bibinfo {year}
  {2006})}\BibitemShut {NoStop}%
\bibitem [{\citenamefont {Chen}\ \emph {et~al.}(2010)\citenamefont {Chen},
  \citenamefont {Gu},\ and\ \citenamefont {Wen}}]{Chen10}%
  \BibitemOpen
  \bibfield  {author} {\bibinfo {author} {\bibfnamefont {X.}~\bibnamefont
  {Chen}}, \bibinfo {author} {\bibfnamefont {Z.-C.}\ \bibnamefont {Gu}}, \ and\
  \bibinfo {author} {\bibfnamefont {X.-G.}\ \bibnamefont {Wen}},\ }\bibfield
  {title} {\enquote {\bibinfo {title} {Local unitary transformation, long-range
  quantum entanglement, wave function renormalization, and topological
  order},}\ }\href {\doibase 10.1103/PhysRevB.82.155138} {\bibfield  {journal}
  {\bibinfo  {journal} {Phys. Rev. B}\ }\textbf {\bibinfo {volume} {82}},\
  \bibinfo {pages} {155138} (\bibinfo {year} {2010})}\BibitemShut {NoStop}%
\bibitem [{\citenamefont {Huang}\ and\ \citenamefont {Chen}(2015)}]{Huang15}%
  \BibitemOpen
  \bibfield  {author} {\bibinfo {author} {\bibfnamefont {Y.}~\bibnamefont
  {Huang}}\ and\ \bibinfo {author} {\bibfnamefont {X.}~\bibnamefont {Chen}},\
  }\bibfield  {title} {\enquote {\bibinfo {title} {Quantum circuit complexity
  of one-dimensional topological phases},}\ }\href {\doibase
  10.1103/PhysRevB.91.195143} {\bibfield  {journal} {\bibinfo  {journal} {Phys.
  Rev. B}\ }\textbf {\bibinfo {volume} {91}},\ \bibinfo {pages} {195143}
  (\bibinfo {year} {2015})}\BibitemShut {NoStop}%
\bibitem [{\citenamefont {Heyl}\ and\ \citenamefont {Budich}(2017)}]{Heyl17}%
  \BibitemOpen
  \bibfield  {author} {\bibinfo {author} {\bibfnamefont {M.}~\bibnamefont
  {Heyl}}\ and\ \bibinfo {author} {\bibfnamefont {J.~C.}\ \bibnamefont
  {Budich}},\ }\bibfield  {title} {\enquote {\bibinfo {title} {Dynamical
  topological quantum phase transitions for mixed states},}\ }\href {\doibase
  10.1103/PhysRevB.96.180304} {\bibfield  {journal} {\bibinfo  {journal} {Phys.
  Rev. B}\ }\textbf {\bibinfo {volume} {96}},\ \bibinfo {pages} {180304}
  (\bibinfo {year} {2017})}\BibitemShut {NoStop}%
\bibitem [{\citenamefont {{D'Alessio}}\ and\ \citenamefont
  {{Rigol}}(2015)}]{Dalessio15}%
  \BibitemOpen
  \bibfield  {author} {\bibinfo {author} {\bibfnamefont {L.}~\bibnamefont
  {{D'Alessio}}}\ and\ \bibinfo {author} {\bibfnamefont {M.}~\bibnamefont
  {{Rigol}}},\ }\bibfield  {title} {\enquote {\bibinfo {title} {{Dynamical
  preparation of Floquet Chern insulators}},}\ }\href {\doibase
  10.1038/ncomms9336} {\bibfield  {journal} {\bibinfo  {journal} {Nat. Commun}\
  }\textbf {\bibinfo {volume} {6}},\ \bibinfo {eid} {8336} (\bibinfo {year}
  {2015})}\BibitemShut {NoStop}%
\bibitem [{\citenamefont {Caio}\ \emph {et~al.}(2015)\citenamefont {Caio},
  \citenamefont {Cooper},\ and\ \citenamefont {Bhaseen}}]{Caio15}%
  \BibitemOpen
  \bibfield  {author} {\bibinfo {author} {\bibfnamefont {M.~D.}\ \bibnamefont
  {Caio}}, \bibinfo {author} {\bibfnamefont {N.~R.}\ \bibnamefont {Cooper}}, \
  and\ \bibinfo {author} {\bibfnamefont {M.~J.}\ \bibnamefont {Bhaseen}},\
  }\bibfield  {title} {\enquote {\bibinfo {title} {{Quantum Quenches in Chern
  Insulators}},}\ }\href {\doibase 10.1103/PhysRevLett.115.236403} {\bibfield
  {journal} {\bibinfo  {journal} {Phys. Rev. Lett.}\ }\textbf {\bibinfo
  {volume} {115}},\ \bibinfo {pages} {236403} (\bibinfo {year}
  {2015})}\BibitemShut {NoStop}%
\bibitem [{\citenamefont {{Vajna}}\ and\ \citenamefont
  {{D{\'o}ra}}(2015)}]{Vajna15}%
  \BibitemOpen
  \bibfield  {author} {\bibinfo {author} {\bibfnamefont {S.}~\bibnamefont
  {{Vajna}}}\ and\ \bibinfo {author} {\bibfnamefont {B.}~\bibnamefont
  {{D{\'o}ra}}},\ }\bibfield  {title} {\enquote {\bibinfo {title} {{Topological
  classification of dynamical phase transitions}},}\ }\href {\doibase
  10.1103/PhysRevB.91.155127} {\bibfield  {journal} {\bibinfo  {journal} {Phys.
  Rev. B}\ }\textbf {\bibinfo {volume} {91}},\ \bibinfo {eid} {155127}
  (\bibinfo {year} {2015})}\BibitemShut {NoStop}%
\bibitem [{\citenamefont {{Wilson}}\ \emph {et~al.}(2016)\citenamefont
  {{Wilson}}, \citenamefont {{Song}},\ and\ \citenamefont
  {{Refael}}}]{Wilson16}%
  \BibitemOpen
  \bibfield  {author} {\bibinfo {author} {\bibfnamefont {J.~H.}\ \bibnamefont
  {{Wilson}}}, \bibinfo {author} {\bibfnamefont {J.~C.~W.}\ \bibnamefont
  {{Song}}}, \ and\ \bibinfo {author} {\bibfnamefont {G.}~\bibnamefont
  {{Refael}}},\ }\bibfield  {title} {\enquote {\bibinfo {title} {{Remnant
  Geometric Hall Response in a Quantum Quench}},}\ }\href {\doibase
  10.1103/PhysRevLett.117.235302} {\bibfield  {journal} {\bibinfo  {journal}
  {Phys. Rev. Lett}\ }\textbf {\bibinfo {volume} {117}},\ \bibinfo {eid}
  {235302} (\bibinfo {year} {2016})}\BibitemShut {NoStop}%
\bibitem [{\citenamefont {Wang}\ \emph {et~al.}(2017)\citenamefont {Wang},
  \citenamefont {Zhang}, \citenamefont {Chen}, \citenamefont {Yu},\ and\
  \citenamefont {Zhai}}]{Wang17}%
  \BibitemOpen
  \bibfield  {author} {\bibinfo {author} {\bibfnamefont {C.}~\bibnamefont
  {Wang}}, \bibinfo {author} {\bibfnamefont {P.}~\bibnamefont {Zhang}},
  \bibinfo {author} {\bibfnamefont {X.}~\bibnamefont {Chen}}, \bibinfo {author}
  {\bibfnamefont {J.}~\bibnamefont {Yu}}, \ and\ \bibinfo {author}
  {\bibfnamefont {H.}~\bibnamefont {Zhai}},\ }\bibfield  {title} {\enquote
  {\bibinfo {title} {{Scheme to Measure the Topological Number of a Chern
  Insulator from Quench Dynamics}},}\ }\href {\doibase
  10.1103/PhysRevLett.118.185701} {\bibfield  {journal} {\bibinfo  {journal}
  {Phys. Rev. Lett.}\ }\textbf {\bibinfo {volume} {118}},\ \bibinfo {pages}
  {185701} (\bibinfo {year} {2017})}\BibitemShut {NoStop}%
\bibitem [{\citenamefont {{Caio}}\ \emph {et~al.}()\citenamefont {{Caio}},
  \citenamefont {{M{\"o}ller}}, \citenamefont {{Cooper}},\ and\ \citenamefont
  {{Bhaseen}}}]{Caio18}%
  \BibitemOpen
  \bibfield  {author} {\bibinfo {author} {\bibfnamefont {M.~D.}\ \bibnamefont
  {{Caio}}}, \bibinfo {author} {\bibfnamefont {G.}~\bibnamefont
  {{M{\"o}ller}}}, \bibinfo {author} {\bibfnamefont {N.~R.}\ \bibnamefont
  {{Cooper}}}, \ and\ \bibinfo {author} {\bibfnamefont {M.~J.}\ \bibnamefont
  {{Bhaseen}}},\ }\bibfield  {title} {\enquote {\bibinfo {title} {{Topological
  Marker Currents in Chern Insulators}},}\ }\href@noop {} {\ }\Eprint
  {http://arxiv.org/abs/1808.10463} {arXiv:1808.10463} \BibitemShut {NoStop}%
\bibitem [{\citenamefont {Heyl}\ \emph {et~al.}(2018)\citenamefont {Heyl},
  \citenamefont {Pollmann},\ and\ \citenamefont {D\'ora}}]{Heyl18}%
  \BibitemOpen
  \bibfield  {author} {\bibinfo {author} {\bibfnamefont {M.}~\bibnamefont
  {Heyl}}, \bibinfo {author} {\bibfnamefont {F.}~\bibnamefont {Pollmann}}, \
  and\ \bibinfo {author} {\bibfnamefont {B.}~\bibnamefont {D\'ora}},\
  }\bibfield  {title} {\enquote {\bibinfo {title} {{Detecting Equilibrium and
  Dynamical Quantum Phase Transitions in Ising Chains via Out-of-Time-Ordered
  Correlators}},}\ }\href {\doibase 10.1103/PhysRevLett.121.016801} {\bibfield
  {journal} {\bibinfo  {journal} {Phys. Rev. Lett.}\ }\textbf {\bibinfo
  {volume} {121}},\ \bibinfo {pages} {016801} (\bibinfo {year}
  {2018})}\BibitemShut {NoStop}%
\bibitem [{\citenamefont {Roy}\ \emph {et~al.}(2017)\citenamefont {Roy},
  \citenamefont {Moessner},\ and\ \citenamefont {Das}}]{Roy17}%
  \BibitemOpen
  \bibfield  {author} {\bibinfo {author} {\bibfnamefont {S.}~\bibnamefont
  {Roy}}, \bibinfo {author} {\bibfnamefont {R.}~\bibnamefont {Moessner}}, \
  and\ \bibinfo {author} {\bibfnamefont {A.}~\bibnamefont {Das}},\ }\bibfield
  {title} {\enquote {\bibinfo {title} {Locating topological phase transitions
  using nonequilibrium signatures in local bulk observables},}\ }\href
  {\doibase 10.1103/PhysRevB.95.041105} {\bibfield  {journal} {\bibinfo
  {journal} {Phys. Rev. B}\ }\textbf {\bibinfo {volume} {95}},\ \bibinfo
  {pages} {041105} (\bibinfo {year} {2017})}\BibitemShut {NoStop}%
\bibitem [{\citenamefont {{Titum}}\ \emph {et~al.}()\citenamefont {{Titum}},
  \citenamefont {{Iosue}}, \citenamefont {{Garrison}}, \citenamefont
  {{Gorshkov}},\ and\ \citenamefont {{Gong}}}]{Paraj18}%
  \BibitemOpen
  \bibfield  {author} {\bibinfo {author} {\bibfnamefont {P.}~\bibnamefont
  {{Titum}}}, \bibinfo {author} {\bibfnamefont {J.~T.}\ \bibnamefont
  {{Iosue}}}, \bibinfo {author} {\bibfnamefont {J.~R.}\ \bibnamefont
  {{Garrison}}}, \bibinfo {author} {\bibfnamefont {A.~V.}\ \bibnamefont
  {{Gorshkov}}}, \ and\ \bibinfo {author} {\bibfnamefont {Z.-X.}\ \bibnamefont
  {{Gong}}},\ }\bibfield  {title} {\enquote {\bibinfo {title} {{Probing
  ground-state phase transitions through quench dynamics}},}\ }\href@noop {} {\
  }\Eprint {http://arxiv.org/abs/1809.06377} {arXiv:1809.06377} \BibitemShut
  {NoStop}%
\bibitem [{\citenamefont {{Zhang}}\ \emph {et~al.}(2017)\citenamefont
  {{Zhang}}, \citenamefont {{Pagano}}, \citenamefont {{Hess}}, \citenamefont
  {{Kyprianidis}}, \citenamefont {{Becker}}, \citenamefont {{Kaplan}},
  \citenamefont {{Gorshkov}}, \citenamefont {{Gong}},\ and\ \citenamefont
  {{Monroe}}}]{Zhang17DPT}%
  \BibitemOpen
  \bibfield  {author} {\bibinfo {author} {\bibfnamefont {J.}~\bibnamefont
  {{Zhang}}}, \bibinfo {author} {\bibfnamefont {G.}~\bibnamefont {{Pagano}}},
  \bibinfo {author} {\bibfnamefont {P.~W.}\ \bibnamefont {{Hess}}}, \bibinfo
  {author} {\bibfnamefont {A.}~\bibnamefont {{Kyprianidis}}}, \bibinfo {author}
  {\bibfnamefont {P.}~\bibnamefont {{Becker}}}, \bibinfo {author}
  {\bibfnamefont {H.}~\bibnamefont {{Kaplan}}}, \bibinfo {author}
  {\bibfnamefont {A.~V.}\ \bibnamefont {{Gorshkov}}}, \bibinfo {author}
  {\bibfnamefont {Z.~X.}\ \bibnamefont {{Gong}}}, \ and\ \bibinfo {author}
  {\bibfnamefont {C.}~\bibnamefont {{Monroe}}},\ }\bibfield  {title} {\enquote
  {\bibinfo {title} {{Observation of a many-body dynamical phase transition
  with a 53-qubit quantum simulator}},}\ }\href {\doibase 10.1038/nature24654}
  {\bibfield  {journal} {\bibinfo  {journal} {Nature}\ }\textbf {\bibinfo
  {volume} {551}},\ \bibinfo {pages} {601--604} (\bibinfo {year}
  {2017})}\BibitemShut {NoStop}%
\bibitem [{\citenamefont {{Jurcevic}}\ \emph {et~al.}(2017)\citenamefont
  {{Jurcevic}}, \citenamefont {{Shen}}, \citenamefont {{Hauke}}, \citenamefont
  {{Maier}}, \citenamefont {{Brydges}}, \citenamefont {{Hempel}}, \citenamefont
  {{Lanyon}}, \citenamefont {{Heyl}}, \citenamefont {{Blatt}},\ and\
  \citenamefont {{Roos}}}]{Jurcevic17}%
  \BibitemOpen
  \bibfield  {author} {\bibinfo {author} {\bibfnamefont {P.}~\bibnamefont
  {{Jurcevic}}}, \bibinfo {author} {\bibfnamefont {H.}~\bibnamefont {{Shen}}},
  \bibinfo {author} {\bibfnamefont {P.}~\bibnamefont {{Hauke}}}, \bibinfo
  {author} {\bibfnamefont {C.}~\bibnamefont {{Maier}}}, \bibinfo {author}
  {\bibfnamefont {T.}~\bibnamefont {{Brydges}}}, \bibinfo {author}
  {\bibfnamefont {C.}~\bibnamefont {{Hempel}}}, \bibinfo {author}
  {\bibfnamefont {B.~P.}\ \bibnamefont {{Lanyon}}}, \bibinfo {author}
  {\bibfnamefont {M.}~\bibnamefont {{Heyl}}}, \bibinfo {author} {\bibfnamefont
  {R.}~\bibnamefont {{Blatt}}}, \ and\ \bibinfo {author} {\bibfnamefont
  {C.~F.}\ \bibnamefont {{Roos}}},\ }\bibfield  {title} {\enquote {\bibinfo
  {title} {{Direct Observation of Dynamical Quantum Phase Transitions in an
  Interacting Many-Body System}},}\ }\href {\doibase
  10.1103/PhysRevLett.119.080501} {\bibfield  {journal} {\bibinfo  {journal}
  {Phys. Rev. Lett.}\ }\textbf {\bibinfo {volume} {119}},\ \bibinfo {eid}
  {080501} (\bibinfo {year} {2017})}\BibitemShut {NoStop}%
\bibitem [{\citenamefont {Fl{\"a}schner}\ \emph {et~al.}(2018)\citenamefont
  {Fl{\"a}schner}, \citenamefont {Vogel}, \citenamefont {Tarnowski},
  \citenamefont {Rem}, \citenamefont {L{\"u}hmann}, \citenamefont {Heyl},
  \citenamefont {Budich}, \citenamefont {Mathey}, \citenamefont {Sengstock},\
  and\ \citenamefont {Weitenberg}}]{Flasch16}%
  \BibitemOpen
  \bibfield  {author} {\bibinfo {author} {\bibfnamefont {N.}~\bibnamefont
  {Fl{\"a}schner}}, \bibinfo {author} {\bibfnamefont {D.}~\bibnamefont
  {Vogel}}, \bibinfo {author} {\bibfnamefont {M.}~\bibnamefont {Tarnowski}},
  \bibinfo {author} {\bibfnamefont {B.~S.}\ \bibnamefont {Rem}}, \bibinfo
  {author} {\bibfnamefont {D.-S.}\ \bibnamefont {L{\"u}hmann}}, \bibinfo
  {author} {\bibfnamefont {M.}~\bibnamefont {Heyl}}, \bibinfo {author}
  {\bibfnamefont {J.C.}\ \bibnamefont {Budich}}, \bibinfo {author}
  {\bibfnamefont {L.}~\bibnamefont {Mathey}}, \bibinfo {author} {\bibfnamefont
  {K.}~\bibnamefont {Sengstock}}, \ and\ \bibinfo {author} {\bibfnamefont
  {C.}~\bibnamefont {Weitenberg}},\ }\bibfield  {title} {\enquote {\bibinfo
  {title} {Observation of dynamical vortices after quenches in a system with
  topology},}\ }\href {https://www.nature.com/articles/s41567-017-0013-8}
  {\bibfield  {journal} {\bibinfo  {journal} {Nat. Phys}\ }\textbf {\bibinfo
  {volume} {14}},\ \bibinfo {pages} {265} (\bibinfo {year} {2018})}\BibitemShut
  {NoStop}%
\bibitem [{\citenamefont {Quan}\ \emph {et~al.}(2006)\citenamefont {Quan},
  \citenamefont {Song}, \citenamefont {Liu}, \citenamefont {Zanardi},\ and\
  \citenamefont {Sun}}]{Quan06}%
  \BibitemOpen
  \bibfield  {author} {\bibinfo {author} {\bibfnamefont {H.~T.}\ \bibnamefont
  {Quan}}, \bibinfo {author} {\bibfnamefont {Z.}~\bibnamefont {Song}}, \bibinfo
  {author} {\bibfnamefont {X.~F.}\ \bibnamefont {Liu}}, \bibinfo {author}
  {\bibfnamefont {P.}~\bibnamefont {Zanardi}}, \ and\ \bibinfo {author}
  {\bibfnamefont {C.~P.}\ \bibnamefont {Sun}},\ }\bibfield  {title} {\enquote
  {\bibinfo {title} {{Decay of Loschmidt Echo Enhanced by Quantum
  Criticality}},}\ }\href {\doibase 10.1103/PhysRevLett.96.140604} {\bibfield
  {journal} {\bibinfo  {journal} {Phys. Rev. Lett.}\ }\textbf {\bibinfo
  {volume} {96}},\ \bibinfo {pages} {140604} (\bibinfo {year}
  {2006})}\BibitemShut {NoStop}%
\bibitem [{\citenamefont {Kells}\ \emph {et~al.}(2014)\citenamefont {Kells},
  \citenamefont {Sen}, \citenamefont {Slingerland},\ and\ \citenamefont
  {Vishveshwara}}]{Kells14}%
  \BibitemOpen
  \bibfield  {author} {\bibinfo {author} {\bibfnamefont {G.}~\bibnamefont
  {Kells}}, \bibinfo {author} {\bibfnamefont {D.}~\bibnamefont {Sen}}, \bibinfo
  {author} {\bibfnamefont {J.~K.}\ \bibnamefont {Slingerland}}, \ and\ \bibinfo
  {author} {\bibfnamefont {S.}~\bibnamefont {Vishveshwara}},\ }\bibfield
  {title} {\enquote {\bibinfo {title} {Topological blocking in quantum quench
  dynamics},}\ }\href {\doibase 10.1103/PhysRevB.89.235130} {\bibfield
  {journal} {\bibinfo  {journal} {Phys. Rev. B}\ }\textbf {\bibinfo {volume}
  {89}},\ \bibinfo {pages} {235130} (\bibinfo {year} {2014})}\BibitemShut
  {NoStop}%
\bibitem [{\citenamefont {{Vodola}}\ \emph {et~al.}(2014)\citenamefont
  {{Vodola}}, \citenamefont {{Lepori}}, \citenamefont {{Ercolessi}},
  \citenamefont {{Gorshkov}},\ and\ \citenamefont {{Pupillo}}}]{Vodola14}%
  \BibitemOpen
  \bibfield  {author} {\bibinfo {author} {\bibfnamefont {D.}~\bibnamefont
  {{Vodola}}}, \bibinfo {author} {\bibfnamefont {L.}~\bibnamefont {{Lepori}}},
  \bibinfo {author} {\bibfnamefont {E.}~\bibnamefont {{Ercolessi}}}, \bibinfo
  {author} {\bibfnamefont {A.~V.}\ \bibnamefont {{Gorshkov}}}, \ and\ \bibinfo
  {author} {\bibfnamefont {G.}~\bibnamefont {{Pupillo}}},\ }\bibfield  {title}
  {\enquote {\bibinfo {title} {{Kitaev Chains with Long-Range Pairing}},}\
  }\href {\doibase 10.1103/PhysRevLett.113.156402} {\bibfield  {journal}
  {\bibinfo  {journal} {Phys. Rev. Lett}\ }\textbf {\bibinfo {volume} {113}},\
  \bibinfo {eid} {156402} (\bibinfo {year} {2014})}\BibitemShut {NoStop}%
\bibitem [{\citenamefont {{Vodola}}\ \emph {et~al.}(2016)\citenamefont
  {{Vodola}}, \citenamefont {{Lepori}}, \citenamefont {{Ercolessi}},\ and\
  \citenamefont {{Pupillo}}}]{Vodola16}%
  \BibitemOpen
  \bibfield  {author} {\bibinfo {author} {\bibfnamefont {D.}~\bibnamefont
  {{Vodola}}}, \bibinfo {author} {\bibfnamefont {L.}~\bibnamefont {{Lepori}}},
  \bibinfo {author} {\bibfnamefont {E.}~\bibnamefont {{Ercolessi}}}, \ and\
  \bibinfo {author} {\bibfnamefont {G.}~\bibnamefont {{Pupillo}}},\ }\bibfield
  {title} {\enquote {\bibinfo {title} {{Long-range Ising and Kitaev models:
  phases, correlations and edge modes}},}\ }\href {\doibase
  10.1088/1367-2630/18/1/015001} {\bibfield  {journal} {\bibinfo  {journal}
  {New J. Phys}\ }\textbf {\bibinfo {volume} {18}},\ \bibinfo {eid} {015001}
  (\bibinfo {year} {2016})}\BibitemShut {NoStop}%
\bibitem [{\citenamefont {{Patrick}}\ \emph {et~al.}(2017)\citenamefont
  {{Patrick}}, \citenamefont {{Neupert}},\ and\ \citenamefont
  {{Pachos}}}]{Patrick17}%
  \BibitemOpen
  \bibfield  {author} {\bibinfo {author} {\bibfnamefont {K.}~\bibnamefont
  {{Patrick}}}, \bibinfo {author} {\bibfnamefont {T.}~\bibnamefont
  {{Neupert}}}, \ and\ \bibinfo {author} {\bibfnamefont {J.~K.}\ \bibnamefont
  {{Pachos}}},\ }\bibfield  {title} {\enquote {\bibinfo {title} {{Topological
  Quantum Liquids with Long-Range Couplings}},}\ }\href {\doibase
  10.1103/PhysRevLett.118.267002} {\bibfield  {journal} {\bibinfo  {journal}
  {Phys. Rev. Lett}\ }\textbf {\bibinfo {volume} {118}},\ \bibinfo {eid}
  {267002} (\bibinfo {year} {2017})}\BibitemShut {NoStop}%
\bibitem [{\citenamefont {Pezz\`e}\ \emph {et~al.}(2017)\citenamefont
  {Pezz\`e}, \citenamefont {Gabbrielli}, \citenamefont {Lepori},\ and\
  \citenamefont {Smerzi}}]{Pezze17}%
  \BibitemOpen
  \bibfield  {author} {\bibinfo {author} {\bibfnamefont {L.}~\bibnamefont
  {Pezz\`e}}, \bibinfo {author} {\bibfnamefont {M.}~\bibnamefont {Gabbrielli}},
  \bibinfo {author} {\bibfnamefont {L.}~\bibnamefont {Lepori}}, \ and\ \bibinfo
  {author} {\bibfnamefont {A.}~\bibnamefont {Smerzi}},\ }\bibfield  {title}
  {\enquote {\bibinfo {title} {Multipartite entanglement in topological quantum
  phases},}\ }\href {\doibase 10.1103/PhysRevLett.119.250401} {\bibfield
  {journal} {\bibinfo  {journal} {Phys. Rev. Lett.}\ }\textbf {\bibinfo
  {volume} {119}},\ \bibinfo {pages} {250401} (\bibinfo {year}
  {2017})}\BibitemShut {NoStop}%
\bibitem [{\citenamefont {{Xiong}}\ \emph {et~al.}()\citenamefont {{Xiong}},
  \citenamefont {{Yao}},\ and\ \citenamefont {{Yan}}}]{Xiong19}%
  \BibitemOpen
  \bibfield  {author} {\bibinfo {author} {\bibfnamefont {Z.}~\bibnamefont
  {{Xiong}}}, \bibinfo {author} {\bibfnamefont {D.-X.}\ \bibnamefont {{Yao}}},
  \ and\ \bibinfo {author} {\bibfnamefont {Z.}~\bibnamefont {{Yan}}},\
  }\bibfield  {title} {\enquote {\bibinfo {title} {{Nonanalyticity of circuit
  complexity across topological phase transitions}},}\ }\href@noop {} {\
  }\Eprint {http://arxiv.org/abs/1906.11279} {arXiv:1906.11279} \BibitemShut
  {NoStop}%
\bibitem [{\citenamefont {Hasan}\ and\ \citenamefont {Kane}(2010)}]{Hasan10}%
  \BibitemOpen
  \bibfield  {author} {\bibinfo {author} {\bibfnamefont {M.~Z.}\ \bibnamefont
  {Hasan}}\ and\ \bibinfo {author} {\bibfnamefont {C.~L.}\ \bibnamefont
  {Kane}},\ }\bibfield  {title} {\enquote {\bibinfo {title} {Colloquium:
  Topological insulators},}\ }\href {\doibase 10.1103/RevModPhys.82.3045}
  {\bibfield  {journal} {\bibinfo  {journal} {Rev. Mod. Phys.}\ }\textbf
  {\bibinfo {volume} {82}},\ \bibinfo {pages} {3045} (\bibinfo {year}
  {2010})}\BibitemShut {NoStop}%
\bibitem [{\citenamefont {Qi}\ and\ \citenamefont {Zhang}(2011)}]{Qi11}%
  \BibitemOpen
  \bibfield  {author} {\bibinfo {author} {\bibfnamefont {X.-L.}\ \bibnamefont
  {Qi}}\ and\ \bibinfo {author} {\bibfnamefont {S.-C.}\ \bibnamefont {Zhang}},\
  }\bibfield  {title} {\enquote {\bibinfo {title} {Topological insulators and
  superconductors},}\ }\href {\doibase 10.1103/RevModPhys.83.1057} {\bibfield
  {journal} {\bibinfo  {journal} {Rev. Mod. Phys.}\ }\textbf {\bibinfo {volume}
  {83}},\ \bibinfo {pages} {1057} (\bibinfo {year} {2011})}\BibitemShut
  {NoStop}%
\bibitem [{\citenamefont {Hyatt}\ \emph {et~al.}(2017)\citenamefont {Hyatt},
  \citenamefont {Garrison},\ and\ \citenamefont {Bauer}}]{Hyatt17}%
  \BibitemOpen
  \bibfield  {author} {\bibinfo {author} {\bibfnamefont {K.}~\bibnamefont
  {Hyatt}}, \bibinfo {author} {\bibfnamefont {J.~R.}\ \bibnamefont {Garrison}},
  \ and\ \bibinfo {author} {\bibfnamefont {B.}~\bibnamefont {Bauer}},\
  }\bibfield  {title} {\enquote {\bibinfo {title} {Extracting entanglement
  geometry from quantum states},}\ }\href {\doibase
  10.1103/PhysRevLett.119.140502} {\bibfield  {journal} {\bibinfo  {journal}
  {Phys. Rev. Lett.}\ }\textbf {\bibinfo {volume} {119}},\ \bibinfo {pages}
  {140502} (\bibinfo {year} {2017})}\BibitemShut {NoStop}%
\bibitem [{\citenamefont {Girolami}(2019)}]{Girolami19}%
  \BibitemOpen
  \bibfield  {author} {\bibinfo {author} {\bibfnamefont {D.}~\bibnamefont
  {Girolami}},\ }\bibfield  {title} {\enquote {\bibinfo {title} {How difficult
  is it to prepare a quantum state?}}\ }\href {\doibase
  10.1103/PhysRevLett.122.010505} {\bibfield  {journal} {\bibinfo  {journal}
  {Phys. Rev. Lett.}\ }\textbf {\bibinfo {volume} {122}},\ \bibinfo {pages}
  {010505} (\bibinfo {year} {2019})}\BibitemShut {NoStop}%
\bibitem [{\citenamefont {Lieb}\ \emph {et~al.}(1961)\citenamefont {Lieb},
  \citenamefont {Schultz},\ and\ \citenamefont {Mattis}}]{Lieb96}%
  \BibitemOpen
  \bibfield  {author} {\bibinfo {author} {\bibfnamefont {E.}~\bibnamefont
  {Lieb}}, \bibinfo {author} {\bibfnamefont {T.}~\bibnamefont {Schultz}}, \
  and\ \bibinfo {author} {\bibfnamefont {D.}~\bibnamefont {Mattis}},\
  }\bibfield  {title} {\enquote {\bibinfo {title} {Two soluble models of an
  antiferromagnetic chain},}\ }\href {\doibase
  https://doi.org/10.1016/0003-4916(61)90115-4} {\bibfield  {journal} {\bibinfo
   {journal} {Ann. Phys.}\ }\textbf {\bibinfo {volume} {16}},\ \bibinfo {pages}
  {407} (\bibinfo {year} {1961})}\BibitemShut {NoStop}%
\bibitem [{\citenamefont {Katsura}(1962)}]{Katsura62}%
  \BibitemOpen
  \bibfield  {author} {\bibinfo {author} {\bibfnamefont {S.}~\bibnamefont
  {Katsura}},\ }\bibfield  {title} {\enquote {\bibinfo {title} {{Statistical
  Mechanics of the Anisotropic Linear Heisenberg Model}},}\ }\href {\doibase
  10.1103/PhysRev.127.1508} {\bibfield  {journal} {\bibinfo  {journal} {Phys.
  Rev.}\ }\textbf {\bibinfo {volume} {127}},\ \bibinfo {pages} {1508} (\bibinfo
  {year} {1962})}\BibitemShut {NoStop}%
\bibitem [{\citenamefont {Perk}\ \emph {et~al.}(1975)\citenamefont {Perk},
  \citenamefont {Capel}, \citenamefont {Zuilhof},\ and\ \citenamefont
  {Siskens}}]{Perk75}%
  \BibitemOpen
  \bibfield  {author} {\bibinfo {author} {\bibfnamefont {J.~H.~H.}\
  \bibnamefont {Perk}}, \bibinfo {author} {\bibfnamefont {H.~W.}\ \bibnamefont
  {Capel}}, \bibinfo {author} {\bibfnamefont {M.~J.}\ \bibnamefont {Zuilhof}},
  \ and\ \bibinfo {author} {\bibfnamefont {Th.~J.}\ \bibnamefont {Siskens}},\
  }\bibfield  {title} {\enquote {\bibinfo {title} {On a soluble model of an
  antiferromagnetic chain with alternating interactions and magnetic
  moments},}\ }\href {\doibase https://doi.org/10.1016/0378-4371(75)90052-7}
  {\bibfield  {journal} {\bibinfo  {journal} {Physica A}\ }\textbf {\bibinfo
  {volume} {81}},\ \bibinfo {pages} {319} (\bibinfo {year} {1975})}\BibitemShut
  {NoStop}%
\bibitem [{\citenamefont {Haldane}(1981)}]{Haldane81}%
  \BibitemOpen
  \bibfield  {author} {\bibinfo {author} {\bibfnamefont {F.~D.~M.}\
  \bibnamefont {Haldane}},\ }\bibfield  {title} {\enquote {\bibinfo {title}
  {{`Luttinger liquid theory'of one-dimensional quantum fluids. I. Properties
  of the Luttinger model and their extension to the general 1D interacting
  spinless Fermi gas}},}\ }\href
  {http://iopscience.iop.org/article/10.1088/0022-3719/14/19/010/meta}
  {\bibfield  {journal} {\bibinfo  {journal} {J. Phys. Condens. Matter}\
  }\textbf {\bibinfo {volume} {14}},\ \bibinfo {pages} {2585} (\bibinfo {year}
  {1981})}\BibitemShut {NoStop}%
\bibitem [{\citenamefont {{Voit}}(1995)}]{Voit95}%
  \BibitemOpen
  \bibfield  {author} {\bibinfo {author} {\bibfnamefont {J.}~\bibnamefont
  {{Voit}}},\ }\bibfield  {title} {\enquote {\bibinfo {title} {{One-dimensional
  Fermi liquids}},}\ }\href {\doibase 10.1088/0034-4885/58/9/002} {\bibfield
  {journal} {\bibinfo  {journal} {Rep. Prog. Phys}\ }\textbf {\bibinfo {volume}
  {58}},\ \bibinfo {pages} {977} (\bibinfo {year} {1995})}\BibitemShut
  {NoStop}%
\bibitem [{\citenamefont {Rahmani}\ and\ \citenamefont
  {Chamon}(2011)}]{Rahman11}%
  \BibitemOpen
  \bibfield  {author} {\bibinfo {author} {\bibfnamefont {A.}~\bibnamefont
  {Rahmani}}\ and\ \bibinfo {author} {\bibfnamefont {C.}~\bibnamefont
  {Chamon}},\ }\bibfield  {title} {\enquote {\bibinfo {title} {{Optimal Control
  for Unitary Preparation of Many-Body States: Application to Luttinger
  Liquids}},}\ }\href {\doibase 10.1103/PhysRevLett.107.016402} {\bibfield
  {journal} {\bibinfo  {journal} {Phys. Rev. Lett.}\ }\textbf {\bibinfo
  {volume} {107}},\ \bibinfo {pages} {016402} (\bibinfo {year}
  {2011})}\BibitemShut {NoStop}%
\bibitem [{\citenamefont {{Ali}}\ \emph {et~al.}()\citenamefont {{Ali}},
  \citenamefont {{Bhattacharyya}}, \citenamefont {{Shajidul Haque}},
  \citenamefont {{Kim}},\ and\ \citenamefont {{Moynihan}}}]{Ali18}%
  \BibitemOpen
  \bibfield  {author} {\bibinfo {author} {\bibfnamefont {T.}~\bibnamefont
  {{Ali}}}, \bibinfo {author} {\bibfnamefont {A.}~\bibnamefont
  {{Bhattacharyya}}}, \bibinfo {author} {\bibfnamefont {S.}~\bibnamefont
  {{Shajidul Haque}}}, \bibinfo {author} {\bibfnamefont {E.~H.}\ \bibnamefont
  {{Kim}}}, \ and\ \bibinfo {author} {\bibfnamefont {N.}~\bibnamefont
  {{Moynihan}}},\ }\bibfield  {title} {\enquote {\bibinfo {title} {{Post-Quench
  Evolution of Distance and Uncertainty in a Topological System: Complexity,
  Entanglement and Revivals}},}\ }\href@noop {} {\ }\Eprint
  {http://arxiv.org/abs/1811.05985} {arXiv:1811.05985} \BibitemShut {NoStop}%
\end{thebibliography}%


\clearpage
\newpage 
\twocolumngrid
\setcounter{figure}{0}
\makeatletter
\renewcommand{\thefigure}{S\@arabic\c@figure}
\setcounter{equation}{0} \makeatletter
\renewcommand \theequation{S\@arabic\c@equation}
\renewcommand \thetable{S\@arabic\c@table}

\begin{center} 
{\large \bf Supplemental Material}
\end{center} 

This Supplemental Material consists of four sections.
In \cref{sec:supp_1}, we analytically derive the circuit complexity for a pair of fermions [Eq.~(\ref{C2}) of the main text].
In \cref{sec:supp_2}, we provide detailed  derivations of the nonanalyticity of circuit complexity, as shown numerically in Figs.~\ref{fig1}(b) and (c) and Figs.~\ref{fig2}(a) and (b) of the main text.  In \cref{sec:supp_3}, we discuss the real-space structure of the optimal circuit. In \cref{sec:supp_4}, we provide numerical and analytical evidence of the nonanalyticity of steady-state circuit complexity after a quantum quench. Finally, in \cref{sec:supp_5}, we provide a detailed analysis of circuit complexity for $p+ip$ topological superconductor.

\section{Derivation of circuit complexity for a pair of fermions}
\label{sec:supp_1}

In this section, we present a detailed derivation of the circuit complexity for a pair of fermions, i.e.\ Eq.~(\ref{C2}) in the main text. This expression has previously been obtained  using different approaches in Refs.~\cite{Hackl18, Khan18, Rey17}. In order to be comprehensive, here we provide a detailed derivation following Ref.~\cite{Khan18}.  We note that Ref.~\cite{Hackl18} provides an alternative derivation using a group theory approach.

By taking the derivative with respect to $s$ in Eq.~(\ref{u_k}) of the main text, we get the following expression:
\begin{equation}
\sum_I Y^I(s)O_I= (\partial_sU(s)) U^{-1} (s),
\end{equation}
where $U(s)$ is a unitary transformation which depends on $s$, and we have omitted the label $k$  for notational clarity. 

The unitary $U(s)$ can be parametrized in matrix form:
\begin{equation}
U(s)= e^{i\beta} \begin{bmatrix} 
    e^{-i \phi_1} \cos{\omega}&\ \  e^{-i \phi_2} \sin{\omega}\\
  -e^{i \phi_2} \sin{\omega} &\ \ e^{i \phi_1} \cos{\omega}
  \end{bmatrix},
\end{equation}
where $\beta, \phi_1, \phi_2, \omega$ explicitly depend  on the parameter $s$. 
The above matrix can be expressed in terms of the  generators of $U(2)$, which we choose as follows:
\begin{equation}
O_0=  \begin{bmatrix} 
    i&0\\
   0&i
  \end{bmatrix},\\ 
  O_1=  \begin{bmatrix} 
    0 & i\\
   i&0
  \end{bmatrix},\\
    O_2=  \begin{bmatrix} 
    0 & 1\\
   -1&0
  \end{bmatrix},\\ 
    O_3=  \begin{bmatrix} 
    i & 0\\
   0&-i
  \end{bmatrix}.
  \label{gen}
\end{equation}
Using the relation 
\begin{equation}
\tr (O_aO_b)= -2 \delta_{ab},
\end{equation}
one can extract the strength, $Y^I(s)$, of generator $O_I$ [cf.\ \cref{u_k} in the main text] as follows:
\begin{equation}
Y^I(s)= -\frac{1}{2} \tr \left[ (\partial_sU(s)) U^{-1} (s) O_I\right ]. 
\end{equation}
Our cost functional can then be expressed as
\begin{eqnarray}
\mathcal{D} &=& \int_0^1 ds \sum_I |Y^I(s)|^2 \nonumber \\
&=& \int_0^1 ds \Bigg[\left(\frac{d \beta}{d s}\right)^2 + \left(\frac{d \omega}{d s}\right)^2 + \cos^2 \omega \left(\frac{d \phi_1}{d s}\right)^2 \nonumber \\
&& \quad \quad \quad \quad + \sin^2 \omega \left(\frac{d \phi_2}{d s}\right)^2\Bigg]. \label{DD}
\end{eqnarray}
Now, by exploiting the boundary condition at $s=0$, i.e.\ $U(s=0)= I$,  we get
\begin{equation}
  \begin{pmatrix}
    \beta(s=0) \\ \phi_1(s=0) \\ \phi_2 (s=0) \\ \omega(s=0)
  \end{pmatrix}
  =
  \begin{pmatrix}
    0 \\ 0 \\ \phi_2(0) \\ 0
    \end{pmatrix},
\end{equation}
where $\phi_2(0)$ is an arbitrary phase.  
Furthermore, we have the boundary condition at $s=1$,
\begin{equation}
U(s=1)= \begin{bmatrix} 
    \cos(\Delta \theta)& - i e^{-i\phi} \sin(\Delta \theta)\\- i  \sin(\Delta \theta)& e^{-i \phi}\cos(\Delta \theta)
    \end{bmatrix},
\end{equation}which results in
\begin{equation}
  \begin{pmatrix}
    \beta(s=1) \\ \phi_1(s=1) \\ \phi_2 (s=1) \\ \omega(s=1)
  \end{pmatrix}
  =
  \begin{pmatrix}
    0 \\ 0 \\ \pi/2 \\ \Delta \theta
  \end{pmatrix}.
\end{equation}

The integrand in Eq.\ (\ref{DD}) is a sum of four non-negative terms. Setting 
$\beta(s) = \phi_1(s) = 0$ and $\phi_2(s) = \pi/2$ 
minimizes (i.e.\ sets to zero) three of the four terms without imposing any additional constraints on the minimization of the remaining $(d \omega/d s)^2$ term. One can then easily check that the linear function 
$w(s) = s \Delta \theta$
minimizes the remaining term and yields
\begin{equation}
\mathcal{C}= \int_0^{1} ds\,|\Delta  \theta|^2 = |\Delta \theta|^2. 
\end{equation}

\section{Analytical Derivation of divergent derivatives in ground states  }
\label{sec:supp_2}

In this section, we provide a detailed analytical derivation to show that the first-order derivative indeed diverges at the critical points in the thermodynamic limit.  We first derive how the derivative diverges when the reference state is in the trivial phase ($|\mu_R|>1$), and then we generalize our results to show how this divergent behavior depends on the particular choice of the reference state. 
Throughout this section we assume the reference lies within a given phase, and allow the target state to approach an arbitrary point in the phase diagram.
Our analytical derivations show that these divergences necessarily map out the  phase boundary, as illustrated in Figs.~\ref{fig2}(a) and (b) in the main text and Fig.~\ref{fig:branchpoints} below. 

We begin with our general expression for the complexity as a function of our reference and target states.
The Bogoliubov angle difference $\Delta \theta_k$ for each momentum sector $k$ can be expressed as
\beq
\resizebox{0.95\hsize}{!}{$\Delta \theta_k = \frac{1}{2} \arctan \frac{\sin k \left[ \Delta_R \mu_T - \Delta_T \mu_R + (\Delta_R - \Delta_T) \cos k \right]}{(\mu_R + \cos k) (\mu_T + \cos k) + \Delta_R \Delta_T \sin^2 k},$}
\label{thetak1}
\eeq
and the circuit complexity is  written in terms of $ \Delta \theta_k$:
\begin{equation}
  \mathcal{C}/L = \frac{1}{2\pi}\int_{0}^{\pi} | \Delta \theta_{k}|^2 \, dk.
  \label{comps}
\end{equation}
Note that we have replaced the discrete sum in the main text with an integral for the thermodynamic limit, and written ``$\mathcal{C}\left( \ket{\Psi_\text{gs}^R} \rightarrow \ket{\Psi_\text{gs}^{T}} \right)$'' as ``$\mathcal{C}$'' for brevity.

Now we substitute \cref{thetak1} into \cref{comps}, and  take the derivatives with respect to $\mu_T$ and $\Delta_T$. We obtain
\bea
\partial_{\mu_T} \mathcal{C}/L &=& \frac{\Delta_T}{4 \pi} \int_{-\pi}^{\pi} \frac{\Delta \theta_k \sin k}{(\mu_T + \cos k)^2 + \Delta_T^2 \sin^2 k} \, dk, \nn
\partial_{\Delta_T} \mathcal{C}/L &=& -\frac{1}{4 \pi} \int_{-\pi}^{\pi} \frac{\Delta \theta_k \sin k \left( \mu_T + \cos k \right)}{(\mu_T + \cos k)^2 + \Delta_T^2 \sin^2 k} \, dk. \qquad
\label{eq:divC}
\eea
Here, we have used the fact that these functions are even in $k$ to extend the integrals to $-\pi$.
In spite of the complicated nature of these integrals, much can be learned about their analytic properties by recasting them as closed contour integrals in the complex plane.
Defining the variable $z = e^{ik}$, we find that the integrals take the form
\bea
\partial_{\mu_T} \mathcal{C}/L &=& - i \Delta_T \oint \frac{dz}{2 \pi i} \frac{ \Delta \theta(z)(z^2 - 1)}{(z^2 + 2 \mu_T z + 1)^2 - \Delta_T^2 (z^2 - 1)^2}, \nn
\partial_{\Delta_T} \mathcal{C}/L &=& \frac{i}{2} \oint \frac{dz}{2 \pi i z} \frac{\Delta \theta(z)\left(z^2 - 1\right)\left( z^2 + 2 \mu_T z + 1 \right)}{(z^2 + 2 \mu_T z + 1)^2 - \Delta_T^2 (z^2 - 1)^2}, \nonumber \\[-.25cm]
&&
\label{eq:divC2}
\eea
where the integration is taken counter-clockwise over the contour $|z| = 1$.
In this form, we may use the fact that the value of the integrals is entirely determined by the non-analyticities of the integrand which are located inside the contour, and that the value of the integration will only diverge if there is a divergence located on the contour.

\begin{figure}
  \centering\includegraphics[height=3.9cm]{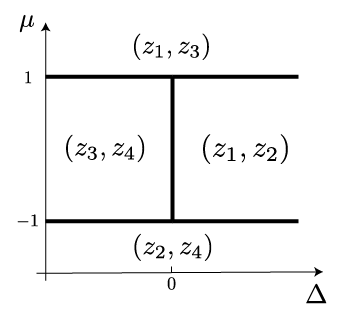}
  \caption{The phase diagram of the Kitaev chain, where in each phase we list which of the two branch points given in Eq.~\eqref{eq:zia} lie inside the contour integrals in Eq.~\eqref{eq:divC2}. The integrals can only diverge at the phase transitions, where the branch points cross the contour, 
  }
  \label{fig:branchpoints}
\end{figure}

We proceed by defining the following variables,
\bea
z_{1,a} &=& \frac{- \mu_a + \sqrt{\mu_a^2 + \Delta_a^2 - 1}}{1 + \Delta_a}, \nn
z_{2,a} &=& \frac{- \mu_a - \sqrt{\mu_a^2 + \Delta_a^2 - 1}}{1 + \Delta_a}, \nn
z_{3,a} &=& \frac{- \mu_a + \sqrt{\mu_a^2 + \Delta_a^2 - 1}}{1 - \Delta_a}, \nn
z_{4,a} &=& \frac{- \mu_a - \sqrt{\mu_a^2 + \Delta_a^2 - 1}}{1 - \Delta_a},
\label{eq:zia}
\eea
where $a = R,T$.
From Eq.~\eqref{eq:divC2}, both derivatives contain simple poles at $z_{i,T}$ for $i=1,2,3,4$, while $\partial_{\Delta_T} \mathcal{C}$ additionally has a simple pole at $z=0$.
Also, using the formula $\arctan(z) = (i/2)\log \frac{1-iz}{1+iz}$, we can write the Bogoliubov angle as
\bea
\Delta\theta(z) &=& \frac{i}{4} \log \Bigg[ \frac{(\Delta_T+1)(z-z_{1,T})(z-z_{2,T})}{(\Delta_T-1)(z-z_{3,T})(z-z_{4,T})} \nn
&& \times \ \frac{(\Delta_R-1)(z-z_{3,R})(z-z_{4,R})}{(\Delta_R+1)(z-z_{1,R})(z-z_{2,R})} \Bigg].
\eea
The important fact we will need is that the complex logarithm contains branch cuts running from the zeros to the infinities of its argument; therefore, the $z_{ia}$ are really branch points of the integrand.
We now note that the derivatives of the complexity will only diverge if the couplings are tuned to a phase transition.
This is because the $z_{i,a}$ can only have unit modulus if we are at one of the phase transitions, and at the phase transitions the branch points cross the contour resulting in a divergent integral, see Fig.~\ref{fig:branchpoints}.
In particular, we may characterize the phase diagram in terms of which branch points are inside or outside the contour integral.

\begin{figure}
  \centering\includegraphics[height=3.9cm]{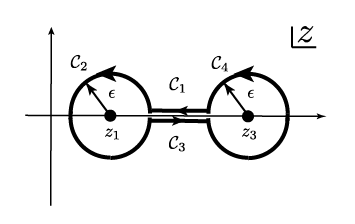}
  \caption{The deformation of the integration contour used to compute the gradients of the circuit complexity in the case $\mu_T > 1$. There is a branch cut running between the branch points $z_1$ and $z_3$, where the imaginary part of the integrand is discontinuous and the integrand diverges near the branch points. 
  }
  \label{fig:intcontour}
\end{figure}

In addition, we may actually compute the integrals exactly in certain cases and limits, which allows us to obtain the exact analytic dependence of the divergence on the couplings.
As a definite example, we consider the case $|\mu_T| > 1$.
In this case, there is a branch cut inside the logarithm running from $z_{1,T}$ to $z_{3,T}$, and one outside between $z_{2,T}$ and $z_{4,T}$, and the divergences seen at $\mu_T \rightarrow 1$ will be due to these branch cuts approaching the contour.
In this case we may entirely factor out the dependence on the reference state from the logarithm and focus on the terms which depend on the target state.
We deform the contour so that it skirts the branch cut [see the parametrization into four contours in Fig.~\ref{fig:intcontour}].
A key point here is that the argument of the logarithm is $-\pi$ upon approaching the branch cut from the bottom-half plane, while it is $+\pi$ upon approaching it from the top half. 
Therefore, in the sum of the two contours running along the branch cut, the logarithm simply contributes a phase factor and we may evaluate the resulting simplified integrand by elementary methods, and for small $\epsilon$ we find
\beq
\resizebox{.95\hsize}{!}{$\int_{\mathcal{C}_1} + \int_{\mathcal{C}_3}  = \frac{1}{16 \sqrt{ \mu_T^2 + \Delta_T^2 - 1}} \log \left| \frac{(z_3 - z_2)(z_1 - z_4) \epsilon^2}{(z_1 - z_2) (z_3 - z_4) (z_1 - z_3)^2} \right|.$}
\eeq
We perform the integral around contour $\mathcal{C}_2$ by writing $z = z_1 + \epsilon e^{i \theta}$, and integrating from $-\pi < \theta < \pi$.
At small $\epsilon$, we find
\beq
\resizebox{.95\hsize}{!}{$\int_{\mathcal{C}_2} = - \frac{1}{16 \sqrt{ \mu_T^2 + \Delta_T^2 - 1}} \log \left| \frac{ (\Delta_T + 1) \epsilon (z_1 - z_2) }{ (\Delta_T - 1) (z_1 - z_3) (z_1 - z_4)} \right|.$}
\eeq
The computation for contour $\mathcal{C}_4$ is similar, although the phase winds around the other way:
\bea
\resizebox{.95\hsize}{!}{$\int_{\mathcal{C}_4} = - \frac{1}{16 \sqrt{ \mu_T^2 + \Delta_T^2 - 1}} \log \left| \frac{ (\Delta_T - 1) \epsilon (z_3 - z_4) }{ (\Delta_T + 1) (z_3 - z_1) (z_3 - z_2)} \right|.$} \nonumber \\
\eea
Finally, taking the sum of all four contours, we find that the $\log \epsilon$ divergence in each integral cancels, and we obtain the desired result:
\bea
\partial_{\mu_T} \mathcal{C}/L &=& \frac{1}{8 \sqrt{ \mu_T^2 + \Delta_T^2 - 1}} \log \left| \frac{\mu_T^2 - 1}{\mu_T^2 + \Delta_T^2 - 1} \right| \nn
&& + \ I_2(\mu_R,\Delta_R,\mu_T,\Delta_T),
\eea
where the function $I_2$ depends on $\mu_{R}$ and $\Delta_R$, but is analytic as the phase transition is approached.
Therefore, when approaching from $\mu_T>1$, the quantity $\partial_{\mu_T} \mathcal{C}/L$ diverges as $\log(\mu_T-1)/8\Delta_T$ if $\Delta_T \neq 0$, but it is analytic if one approaches the multicritical point at $\Delta_T = 0$.

Similar manipulations may be made for $\partial_{\Delta_T} \mathcal{C}/L$ and in other phases.
Sometimes the branch cuts take a complicated form in the complex plane so that we cannot reduce the expression into elementary integrals, but we can still deduce the form of the divergence by considering how the contour integrals behave as the branch points cross the contour.

Our final results are summarized as follows.
The expression $\partial_{\mu_T} \mathcal{C}/L$ is always analytic unless $\mu_T \rightarrow \pm1$.
Near these phase transitions, it diverges as
\beq
\partial_{\mu_T} \mathcal{C}/L \sim \frac{\mathrm{sign}(\mu_T)}{8 \sqrt{ \mu_T^2 + \Delta_T^2 - 1}} \log \left| \frac{\mu_T^2 - 1}{\mu_T^2 + \Delta_T^2 - 1} \right|,
\eeq
so the divergence is $\mathrm{sign}(\mu_T)\log|\mu_T-1|/8\Delta_T$ if $\Delta_T \neq 0$, but there is not a divergence at $\Delta_T = 0$.

In contrast, the expression $\partial_{\Delta_T} \mathcal{C}/L$ is analytic whenever $\Delta_T \neq 0$.
In this case, the divergence depends on whether the couplings $(\mu_T,\Delta_T)$ approach the phase transitions from the topological phase or the trivial phase.
If we approach the multicritical points from the trivial phases, we find that $\partial_{\Delta_T} \mathcal{C}/L$ remains analytic.  
In contrast, if we approach $\Delta_T = 0$ from the topological phases, we find
\beq
\partial_{\Delta_T} \mathcal{C}/L \sim \frac{1}{4} \left( 1 + \frac{|\mu_T \Delta_T|}{\sqrt{|\mu_T^2 + \Delta_T^2 - 1|}} \right)\log \left| \Delta_T\right|.
\eeq
In this case, we have a $\log|\Delta_T|/4$ divergence when $|\mu_T| < 1$, but now we find that the divergence crosses over to $\log|\Delta_T|/2$ as we approach the multicritical points.

\section{Real-space behavior of the optimal circuits}
\label{sec:supp_3}

In this section, we show how that the real-space optimal circuit behaves differently depending on whether or not the initial and target states are in the same topological phase.

As we have derived in \cref{sec:supp_1} of the Supplemental Material, for a single momentum sector $k$, the circuit complexity is found to be the squared difference between the Bogoliubov angles [Eq.\ (\ref{C2}) in the main text], and the optimal circuit is generated by the following time-independent Hamiltonian,
\begin{equation}
\mathcal{H}_k = - \Delta \theta_k \, O_{1,k}, 
\end{equation}
where $O_{1,k}$ is the same generator given by Eq.\ (\ref{gen}) for momentum sector $k$. Here, we have omitted the time label `$s$' for simplicity as the circuit is time independent (and the total evolution time is fixed to be constant 1). 
As in the main text and following the circuit complexity literature, we have defined $\mathcal{H}_k$ to be anti-Hermitian [\cref{u_k}].

Since the ground state of the Hamiltonian is a product of all momentum sectors with $k>0$, the optimal circuit which generates the evolution between two ground states can be written as
\begin{equation}
\mathcal{H} = \sum_{k>0} \mathcal{H}_k  = \sum_{k>0}-\Delta \theta_{k}  \, O_{1, k}. 
\end{equation}
We are interested in the real-space behavior of the above Hamiltonian.
To discern this, we first write the above Hamiltonian in operator form
\begin{equation}
\mathcal{H} = \sum_{k>0} \mathcal{H}_k  = \sum_{k>0} - i \Delta \theta(k) \hat{\psi}_k^\dagger \tau_1 \hat{\psi}_k^{\phantom\dagger},
\end{equation}
where $\tau_i$ are the Pauli matrices, and $\hat{\psi}_k$ denotes the Nambu spinor
\begin{equation}
    \hat{\psi}_k = \left(\begin{array}{c}
    \hat{a}_k \\
    \hat{a}_{-k}^\dagger \\
    \end{array}\right).
\end{equation}
Utilizing the particle-hole symmetry of the Nambu spinor 
\begin{equation}
    \hat{\psi}_{-k} = \tau_1 (\hat{\psi}_k^\dagger)^T,
\end{equation}
we can extend the sum in the evolution Hamiltonian to be over the entire Brillouin zone
\begin{equation}
   \mathcal{H}= \sum_k-i  \omega(k) \hat{\psi}_k^\dagger \tau_1 \hat{\psi}_k^{\phantom\dagger},
\end{equation}
where $\omega(k)$ satisfies 
\begin{equation}
    \omega(k) -\omega(-k) = \Delta \theta(k)
\end{equation}
for $k>0$. 
In particular, only the odd part of the function contributes since the even part cancels in the $\tau_1$ pairing channel.

We now proceed by performing a Fourier series expansion of the function $\omega(k)$ over the Brillouin zone.
Without loss of generality we may consider only the odd Fourier series since the even terms will cancel. 
Thus, we write 
\begin{equation}
    \omega(k) = \sum_{n=1}^{\infty} \omega_n \sin(nk) =  \frac{\Delta \theta(k)}{2},
\end{equation}
where the last equality is used to determine the Fourier coefficients. 

Our crucial observation is that when the two states are within the same phase, the Fourier sine series for $\Delta \theta(k)$ ought to be {\it uniformly convergent}.  
This can be seen by considering the boundary conditions, which in this case read $\Delta \theta(0) = \Delta\theta(\pi) = 0$, as shown in Fig.~\ref{fig1}(d) in the main text.  
Thus, if we allow the time-evolved state $\ket{\Psi_T^{\prime}}$ to be within an arbitrarily small error $\epsilon$ to the real target state $\ket{\Psi_T}$, this Fourier series can be accurately truncated to a finite order $N^*$ over the entire Brillouin zone. 

This is relevant because in real-space, the Fourier harmonic $\sin (l k) \,   \hat{\psi}_k^{\dagger} \tau_1 \hat{\psi}_k^{\phantom\dagger}$ is generated by a term involving two fermionic operators separated by $l$ sites. 
More specifically, as this occurs in the $\tau_1$ channel, $\mathcal{H} $ must involve real-space pairing terms such that 
\begin{equation}
    \mathcal{H} = \sum_{j}\sum_{n=1}^{N^*}   \omega_n  \left( \hat{a}_j \hat{a}_{j+n}  -\textrm{H.c.} \right).
\end{equation}
The above argument holds when the system size $L$ is taken to be infinite. In such a case, the finite-range interacting evolution Hamiltonian can be regarded as a truly short-range Hamiltonian, and our results imply that the optimal circuit (with constant time or depth) which evolve states within the same phase region is {\it short-range}. 

On the other hand, when the two states are in different phases, the boundary conditions $\Delta \theta(\pi) = \pi/2 \neq \Delta\theta(0) = 0$ obstruct uniform convergence, analogous to the Gibbs phenomenon.
In this case, the Fourier sine series may still converge pointwise, but for fixed error the series {\it cannot} be truncated to finite order $N^*$ over the entire Brillouin zone.
In such cases, the optimal evolution Hamiltonian $\mathcal{H}$ that transforms states between different topological phases must be {\it long-range} when the evolution time is fixed to be a constant.
Again, this is because the longest real-space distance required to generate the evolution Hamiltonian is given by the highest order of Fourier mode appearing in the momentum space series, which now cannot be accurately truncated.

\section{Numerical evidence for nonanalyticity of quench dynamics  }
\label{sec:supp_4}

In this section, we provide detailed numerical explanations for the nonanalyticity of the long-time steady-state value of the circuit complexity at critical points, as observed in Fig.~\ref{fig3}(b) of the main text.  

As derived in the main text, the time-dependent circuit complexity is given by
 \begin{equation}
\mathcal{C}(\ket{\Psi_i} \rightarrow \ket{\Psi(t)}) =\sum_{k_n} \phi_{k_n}^2  (t),
\end{equation}
where
\begin{equation} \label{phi_kn_supp}
  \phi_{k_n}  (t)= \arccos\sqrt{1-\sin^2(2\Delta \theta_{k_n}) \sin^2(\varepsilon_{k_n}t)}.
\end{equation}

Then the long-time steady-state complexity is just given by  the time-averaged value of the above expression,
\begin{equation}
\overline{\mathcal{C}(\ket{\Psi_i} \rightarrow \ket{\Psi(t)}) }=\sum_{k_n} \overline{ \phi_{k_n}^2  (t)},
\end{equation}
where the overline denotes time averaging.
Because ${ \phi_{k_n}^2  (t)}$
is such a complex expression, it is unknown to us how to derive an analytical function for the time-averaged circuit complexity. Instead, we plot ${\overline{ \phi_{k_n}  (t)}}$ numerically, and show that the nonanalyticity indeed occurs at the phase transition.

\begin{figure}
  \centering\includegraphics[width=0.498\textwidth]{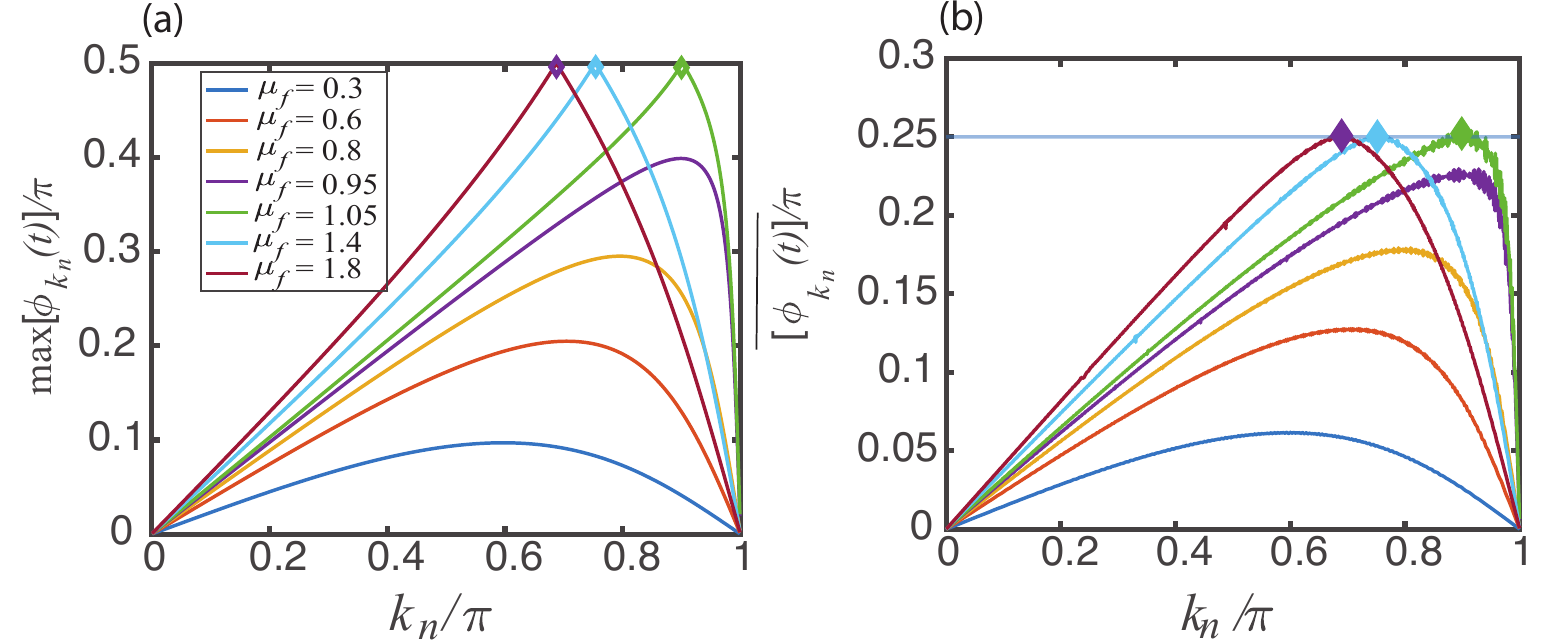}
  \caption{(a) Maximum value of $\phi_{k_n}(t)$ versus $k_n$ for different post-quench Hamiltonian parameters. (b)  Time-averaged value of $\phi_{k_n}(t)$ versus $k_n$ for different post-quench Hamiltonian parameters.  In both panels, $\mu_i=0$, $\Delta_{i}=\Delta_{f}= 1$, and $L=1000$. The diamond markers denote the expected locations of the maxima across the phase transition, given by solutions to $1+\mu_f \cos{k_n}=0$ (see text). }
  \label{fig_s1}
\end{figure}

From the expression of $\phi_{k_n}(t)$, it is clear  that its value oscillates with time, and it reaches its maximal value (envelope) for each momentum sector $k_n$ when $\sin(\epsilon_{k_n}t)=1$.  In Fig.~\ref{fig_s1}(a), we plot the maximum value of $\phi_{k_n}(t)$ for different post-quench Hamiltonian parameters. As the figure clearly shows, when the chemical potential $\mu_f$ of the post-quench Hamiltonian is below the critical value ($\mu_f=1$), $\max [\phi_{k_n}(t)]$ is a smooth function of $k_n$. However, when $\mu_f$ is above the critical value, $\max [\phi_{k_n}(t)]$ exhibits a kink at a certain momentum $k_n$, with its maximal value reaching  $\frac{\pi}{2}$.  To understand this behavior, we can write down the expression for $\max [\phi_{k_n}(t)]$ given the choice of parameters $\mu_i=0, \Delta_{i}=\Delta_{f}=1$:
\begin{equation}
\max [\phi_{k_n}(t)]= \arccos {\left|\frac{1+\mu_f \cos{k_n}}{\sqrt{\mu_f^2+ 2\mu_f \cos{k_n} +1}}\right|}.
\end{equation}
From the above expression, it is clear that when $\mu_f<1$,  $\max [\phi_{k_n}(t)]$  is always smaller than $\pi/2$; when $\mu_f>1$, $\max [\phi_{k_n}(t)]$ can obtain the maximal value of $\pi/2$ when $1+\mu_f \cos{k_n}=0$.  Because one needs to take the absolute value for the arguments of $\arccos$, the quantity $\max [\phi_{k_n}(t)]$   exhibits a kink when reaching $\pi/2$, in agreement with Fig.~\ref{fig_s1}(a). 

We plot the time-averaged value of $\phi_{k_n}(t)$ in Fig.~\ref{fig_s1}(b). Again, we see an upper bound of $\overline{\phi_{k_n}(t)}$ when quenching across the critical point.  Similar to Fig.~\ref{fig_s1}(a), $\overline{\phi_{k_n}(t)}$ reaches its maximal value when $1+\mu_f \cos{k_n}=0$, i.e.\ when $\sin(2\Delta \theta_{k_n})=1$.  For this special momentum sector, the expression for $\phi_{k_n}(t)$ can be written as
\begin{equation}
\phi_{k_n}  (t)= \arcsin{ \left|\sin(\varepsilon_{k_n}t)\right|}.
\end{equation}
Clearly, the time-averaged value of the above expression is just $\pi/4$, in agreement with the numerical results shown in Fig.~\ref{fig_s1}(b).  Therefore, after the phase transition takes place, the maximal value of $\overline{\phi_{k_n}(t)}$ is bounded by $\pi/4$. (This feature is independent of the parameters of the pre-quench Hamiltonian.) 

Having revealed this feature of $\overline{\phi_{k_n}(t)}$, the nonanalyticity  can be understood as follows: as $\mu_f$ increases but is still below the phase transition point, the integral of $\overline{\phi_{k_n}^2(t)}$ increases smoothly with $\mu_f$. After reaching the phase transition, $\overline{\phi_{k_n}^2(t)}$ saturates the bound, and thus the integral's (circuit complexity's) dependence on $\mu_f$ takes a different form. In particular, for the parameters shown in Fig.~\ref{fig_s1} [blue line in Fig.~\ref{fig3}(b) in the main text], the integral (i.e., the circuit complexity) becomes  a constant after the phase transition. This leads to a clear nonanalytical (kink) point at $\mu_f=1$. 

\section{Circuit complexity for two-dimensional $p+ip$ topological superconductors}
\label{sec:supp_5}

In this section, we show how our results for the 1D Kitaev chain can be generalized to 2D\@. In particular, we consider a $p+ip$ topological superconductor for which the Hamiltonian can be written in momentum space as:
\begin{equation}
\hat{H} = \sum_{\bm k} \hat{\psi}_{\bm k}^\dagger \mathcal{H}_{\bm k} \hat{\psi}_{\bm k},
\end{equation}
where the summation is taken over the 2D Brillouin zone, and $\hat{\psi}_k = \left(\begin{array}{c}
    \hat{a}_k \\
    \hat{a}_{-k}^\dagger \\
    \end{array}\right)$ is the Nambu spinor. The single-particle
    Hamiltonian takes the following form:
\begin{equation}
    \mathcal{H}_{\bm k} = 
    \begin{pmatrix}
    \varepsilon_{\bm k} & \Delta_{\bm k}^*  \\
    \Delta_{\bm k}  & -\varepsilon_{\bm k}
    \end{pmatrix},
\end{equation}
where $\varepsilon_{\bm k}$ and $\Delta_{\bm k}$ denote the kinetic and pairing terms in 2D respectively. The ground state wavefunction can be written as
\begin{equation}
\ket{\Psi_{\text{gs}}}= \prod_{\bm k} (\cos \theta_{\bm k} - i\sin \theta_{\bm k} \hat{a}_{\bm k}^{\dagger}\hat{a}_{-\bm k}^{\dagger}) \ket{0},
\end{equation}
where ${\tan}(2\theta_{\bm k}) = |\Delta_{\bm k}|/\varepsilon_{\bm k}$. Similar to 1D, the circuit complexity of the full wavefunction is given by
\begin{equation}
    \mathcal{C} = \sum_{\bm k} |\Delta \theta_{\bm k}|^2 = \frac{L^2}{(2\pi)^2}\int {\rm d}^2{\bm k}|\Delta \theta({\bm k})|^2,
\end{equation}
where we have replaced the summmation by an integral in the infinite-system limit. In this continuum limit, $\varepsilon({\bm k}) \approx \frac{k^2}{2m} - \mu$ and $\Delta({\bm k}) \approx i \Delta (k_x + ik_y)$.

We expect that the non-analyticity should not depend on the particular choice of initial reference state, so we take $\mu_R \rightarrow -\infty$ [with $\theta^R({\bm k})=0$] for simplicity. This corresponds to the trivial vacuum with no particle. Upon tuning $\mu$, the system undergoes a quantum phase transition into the topological phase at $\mu=0$. Taking the derivative of $\mathcal{C}$ with respect to $\mu_T$, we obtain
\begin{eqnarray}
\partial_{\mu_T} \mathcal{C} &=& \frac{L^2}{(2\pi)^2} \int {\rm d}^2 {\bm k} \ 2 \theta^T({\bm k}) \partial_{\mu_T} \theta^T({\bm k})  \nonumber  \\
&=& \frac{L^2}{(2\pi)^2} \int {\rm d}^2 {\bm k} \theta^T({\bm k}) \partial_{\mu_T} \left[ {\arctan} \frac{|\Delta({\bm k})|}{\varepsilon({\bm k})}\right]  \nonumber \\
&=& \frac{L^2}{(2\pi)^2} \int {\rm d}^2 {\bm k} \frac{\theta^T({\bm k}) |\Delta({\bm k})|}{E({\bm k})^2}.
\end{eqnarray}

\end{document}